\documentclass[12pt]{article}
\usepackage{authblk}

\usepackage{mathtools}
\usepackage{amsmath,amsthm,amssymb}
\usepackage{bm}
\usepackage{listings}
\usepackage{natbib} 
\usepackage{float}
\usepackage{mathtools}
\usepackage{ wasysym }

\usepackage[nodisplayskipstretch]{setspace}
\usepackage[dvipsnames]{xcolor}
\usepackage{graphicx}
\usepackage{subcaption}
\usepackage{mwe}
\usepackage[toc,page]{appendix}
\usepackage[nodisplayskipstretch]{setspace}
\usepackage{booktabs}
\usepackage{url}
\usepackage{lscape}
\makeatletter
\g@addto@macro{\UrlBreaks}{\UrlOrds}
\makeatother
\usepackage{enumerate}

\usepackage{pdfpages}

\usepackage{hyperref}

\newcommand\myshade{95}
\colorlet{mylinkcolor}{violet}
\colorlet{mycitecolor}{blue}
\colorlet{myurlcolor}{MidnightBlue}

\hypersetup{
  linkcolor  = mylinkcolor!\myshade!black,
  citecolor  = mycitecolor!\myshade!black,
  urlcolor   = myurlcolor!\myshade!black,
  colorlinks = true,
}

\newcommand{\E}{\mathbb{E}}
\newcommand{\V}{\mathbb{V}}

\newcommand{\Dcal}{\mathcal{D}}

\newcommand{\bfX}{{\mathbf{X}}}
\newcommand{\bfY}{{\mathbf{Y}}}
\newcommand{\bfZ}{{\mathbf{Z}}}
\newcommand{\bfx}{{\mathbf{x}}}
\newcommand{\bfy}{{\mathbf{y}}}
\newcommand{\bfz}{{\mathbf{z}}}

\newcommand{\bfw}{{\mathbf{w}}}
\newcommand{\bfc}{{\mathbf{c}}}
\newcommand{\bfb}{{\mathbf{b}}}
\newcommand{\bmeps}{{\bm{\varepsilon}}}

\newcommand{\bmeta}{{\bm{\eta}}}

\newcommand{\bmtau}{{\bm{\tau}}}

\newcommand{\bmx}{{\bm{x}}}
\newcommand{\bmy}{{\bm{y}}}
\newcommand{\bmz}{{\bm{z}}}

\newcommand{\bmc}{{\bm{c}}}

\newcommand{\ttP}{{\mathtt{P}}}
\newcommand{\ttQ}{{\mathtt{Q}}}
\newcommand{\ttR}{{\mathtt{R}}}

\newcommand{\p}{\mathsf{p}}

\newcommand{\Real}{\mathbb{R}}

\newcommand{\e}{\textsf{e}}

\newcommand{\inv}{^{-1}}

\renewcommand{\exp}{\mathsf{exp}}
\renewcommand{\log}{\,\mathsf{log}\,}

\renewcommand{\min}{\mathsf{min}}
\renewcommand{\max}{\mathsf{max}}

\newcommand{\diag}{\mathsf{diag}}

\newcommand{\Ind}[1]{\mathbb{I}_{\left\{#1\right\}}}    
\newcommand{\Id}[1]{{I}_{#1}}

\newcommand{\OVec}{\mathbf{0}}                          

\newcommand{\Th}{\textsuperscript{th}~} 

\newcommand{\normal}{\mathsf{N}}

\newcommand{\unif}{\mathsf{Unif}}
\newcommand{\gamdist}{\mathsf{Gamma}}
\newcommand{\betadist}{\mathsf{Beta}}
\newcommand{\Exp}{\mathsf{Exp}}


\theoremstyle{definition}

\AtBeginEnvironment{subappendices}{%
\chapter*{Appendices}
\addcontentsline{toc}{chapter}{Appendices}
\counterwithin{figure}{section}
\counterwithin{table}{section}
}

\AtEndEnvironment{subappendices}{%
\counterwithout{figure}{section}
\counterwithout{table}{section}
}

\begin{document}

\title{Predicting milk traits from spectral data using Bayesian probabilistic partial least squares regression} %
\author[1]{Szymon Urbas\thanks{szymon.urbas@ucd.ie}}
\author[2]{Pierre Lovera}
\author[2]{Robert Daly}
\author[2]{Alan O'Riordan}
\author[3]{Donagh Berry} 
\author[1]{Isobel Claire Gormley} 

\affil[1]{\footnotesize{School of Mathematics \& Statistics,  University College Dublin.}}
\affil[2]{\footnotesize{Tyndall National Institute, University College Cork.}}
\affil[3]{\footnotesize{Animal \& Grassland Research and Innovation Centre, Teagasc.}}

\date{}

\maketitle
\begin{abstract}
High-dimensional spectral data---routinely generated in dairy production---are used to predict a range of traits in milk products. Partial least squares (PLS) regression is ubiquitously used for these prediction tasks. However, PLS regression is not typically viewed as arising from a probabilistic model, and parameter uncertainty is rarely quantified. Additionally, PLS regression does not easily lend itself to model-based modifications, coherent prediction intervals are not readily available, and the process of choosing the latent-space dimension, $\mathtt{Q}$, can be subjective and sensitive to data size.
 
We introduce a Bayesian latent-variable model, emulating the desirable properties of PLS regression while accounting for parameter uncertainty in prediction. The need to choose $\mathtt{Q}$ is eschewed through a nonparametric shrinkage prior. The flexibility of the proposed Bayesian partial least squares (BPLS) regression framework is exemplified by considering sparsity modifications and allowing for multivariate response prediction.

The BPLS regression framework is used in two motivating settings: 1) multivariate trait prediction from mid-infrared spectral analyses of milk samples, and 2) milk pH prediction from surface-enhanced Raman spectral data. The prediction performance of BPLS regression at least matches that of PLS regression. Additionally, the provision of correctly calibrated prediction intervals objectively provides richer, more informative inference for stakeholders in dairy production.

\end{abstract}
\section{Introduction}
Quality traits of animal-based products are important determinants of not only the nutritive value of the resulting product but also provide information on the optimal product processing strategy or product portfolio from the raw product. A rapid, low-cost, accurate and non-destructible method to quantify multiple quality parameters from a single sample, ideally in real time, is a requirement of the agri-food sector. Individual animal or bulk milk samples---taken globally from a range of different species---are subjected to spectroscopy analysis, usually in the mid-infrared region of the electromagnetic spectrum. Many studies have demonstrated how the generated spectrum can be used to predict milk quality \citep[e.g.][]{DeTo2014} and other animal level attributes including energy status \citep{FrGo2023} and methane emissions \citep{CoVa2022}. The advantage of such a strategy is that the developed prediction model can be rapidly deployed to routinely collected samples to phenotype both the individual and bulk samples at little to no additional cost.

Partial least squares (PLS) regression is the industry-standard approach to developing prediction models using high-dimensional spectral data. This approach is also popular in the field of chemometrics \citep[e.g.][]{FrFr1993,Kour2002} and in other areas such as econometrics \citep{NaTs2000} and genomics \citep{BoSt2007}.  Partial least squares offers solutions to  \emph{``small $N$, large $\mathtt{P}$''} prediction problems, where the number of observations in the training (calibration) data, $N$, is limited relative to the data dimensionality, $\mathtt{P}$, and the predictor variables exhibit strong correlations.

The partial least squares (PLS) regression method quantifies the relationship between continuous predictor and response variables through a linear projection onto a lower-dimensional latent space. The particular projection is obtained by maximising the norm of the observed cross-covariance between the variables through algorithms such as NIPALS \citep{Wold1973} or SIMPLS \citep{deJo1993}. 

Over the years, variants of PLS emerged to improve predictions under different settings; these include sparse PLS \citep[sPLS,][]{ChKe2010}, robust PLS \citep{HuBr2003}, or kernel-based approaches \citep[e.g.][]{LiGe1993}. 
An intrinsic limitation of PLS regression is that none of the optimisation algorithms arise as a solution to a particular inference problem; this is in contrast to principal component analysis \citep[PCA, e.g.~Chapter 12 of][]{BiNa2006}, a closely related dimension-reduction method, which was shown to be the maximum likelihood solution to a specific Gaussian factor model \citep{TiBi1999}. To address this,  approaches that emulate the PLS predictions through an appropriate probabilistic model have emerged; examples include \cite{LiGa2010}, \cite{Viva2013}, \cite{ZhSo2016}, \cite{BoUh2018} and \cite{BoUh2022}. Despite their coherent probabilistic frameworks, these prediction models focused on point predictions but did not consider their associated uncertainty, either through the assumed data-generating mechanism or as introduced by parameter uncertainty. Additionally, whilst in principle such methods could model a multivariate response, in practice they have limitations, e.g.\  frequentist inference of probabilistic formulations of PLS requires limiting constraints on different parts of the models. Moreover, PLS regression requires the specification of the number of latent variables which is usually informed through cross-validation based on the predictive error of a response variable---it is often more practical to fit multiple univariate models, leading to less coherent and interpretable regression.


Motivated by the need for coherent, richer inference when simultaneously predicting multiple milk traits from spectral data, we introduce Bayesian partial least squares (BPLS) regression, a latent-variable Gaussian model, which emulates PLS regression and addresses its limitations with regard to inference. The model inference is conducted in the Bayesian paradigm thereby elegantly accounting for uncertainty. A nonparametric prior induces shrinkage, alleviating the difficult and often subjective choice of the latent dimension, particularly in settings involving multivariate traits. Finally, we exploit the model-based framework of BPLS by facilitating multivariate trait prediction and consider sparsity modifications providing increased model flexibility. Collectively these aspects of BPLS lead to a user-friendly and statistically-principled prediction method, applicable to a wide range of problems in the agri-food production sector. We demonstrate the performance of BPLS regression through synthetic examples prior to its applications to the real-life problem of predicting milk phenotypes from spectral data. Resulting point predictions are at least as accurate as those produced by the industry-standard PLS regression, with the provision of correctly calibrated prediction intervals objectively providing richer and more informative inference.

In what follows, Section \ref{sec:data_descriptions} introduces the mid-infrared and surface-enhanced Raman spectral datasets from the Irish agri-food sector which motivate this study. Section \ref{sec:model} describes the BPLS regression model, its sparsity-inducing modifications and its associated inference. Section \ref{sec:numerical_experiments} demonstrates the performance of BPLS regression on synthetic and benchmark data sets while Section \ref{sec:data_examples} explores the results of its motivating application to predict a range of milk traits from spectral data. Section \ref{sec:discussion} concludes with a discussion of the BPLS contribution and a delineation of some potential future research avenues.

The BPLS models and inference were implemented in \textsc{R}, and the code with which all results were generated is freely
available at: \url{https://github.com/SzymonUrbas/bplsr}.

\section{Milk spectral data from the Irish agri-food sector} 
\label{sec:data_descriptions}

While having general applicability, the BPLS methods introduced here are motivated by the problem of predicting a range of milk traits from spectral data as encountered in the Irish agri-food sector, whilst accounting for the inherent prediction uncertainty. Two motivating spectral datasets are considered: the first considers the prediction of various chemical and technological milk traits from mid-infrared (MIR) spectra, and the second focuses on predicting the pH of a milk sample using surface-enhanced Raman spectroscopy (SERS) data. In MIR data, the predictors are based on spectra of absorbances given as $\log(1/R_i)$, where $R_i$ is the reflectance of the $i^{\mathrm{th}}$ wavenumber in a given dataset; the reader is referred to \cite{BePe1998} for further details. In SERS, the predictors are logarithms of intensities of light scattering at different wavenumbers. 

Throughout, we denote a predictor vector by $\bfx\in\Real^\mathtt{P}$, where $\mathtt{P}$ is the number of wavenumbers, and denote by $\bfy\in\Real^\mathtt{R}$ a vector of $\mathtt{R}$ responses or traits, noting that BPLS regression can consider multiple traits simultaneously, i.e.~$\mathtt{R} \ge 1$. Each dataset has $N$ samples and $\bfX$ denotes the $N\times\mathtt{P}$ matrix whose $n^{\mathrm{th}}$ row is $\bfx_n^\top$, with a similar definition for $\bfY\in\Real^{N\times\mathtt{R}}$. 

In what follows, all variables are standardised before inference. Additionally, cognisant of the practical utility of the BPLS predictions, traits $\bfY$ which are strictly positive are transformed as outlined in Appendix \ref{app:resp_transform} of the Supplementary Material.


\begin{figure}[t]
\centering
         \includegraphics[width=0.49\textwidth, height = 5cm]{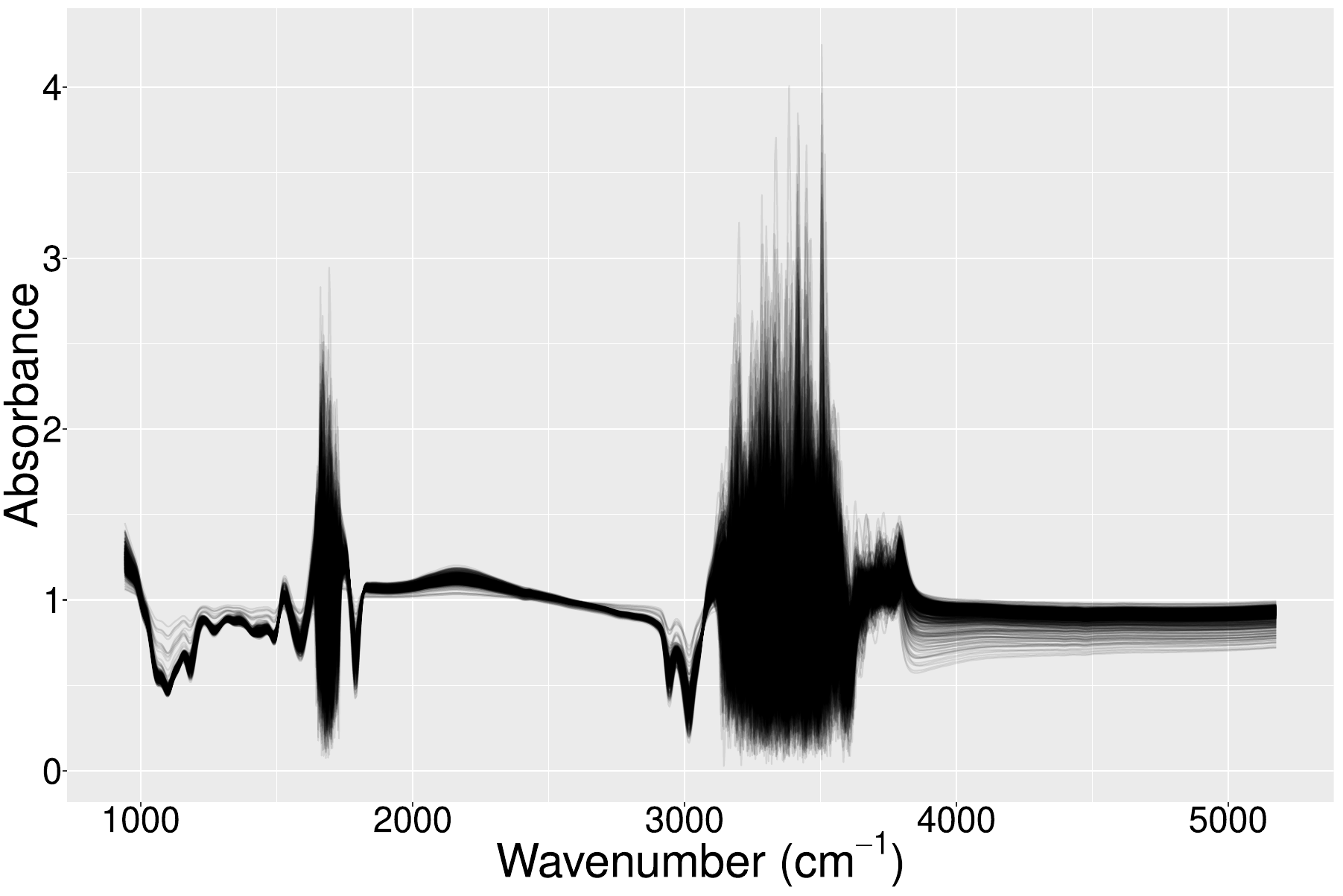}
         \includegraphics[width=0.49\textwidth, height = 5cm]{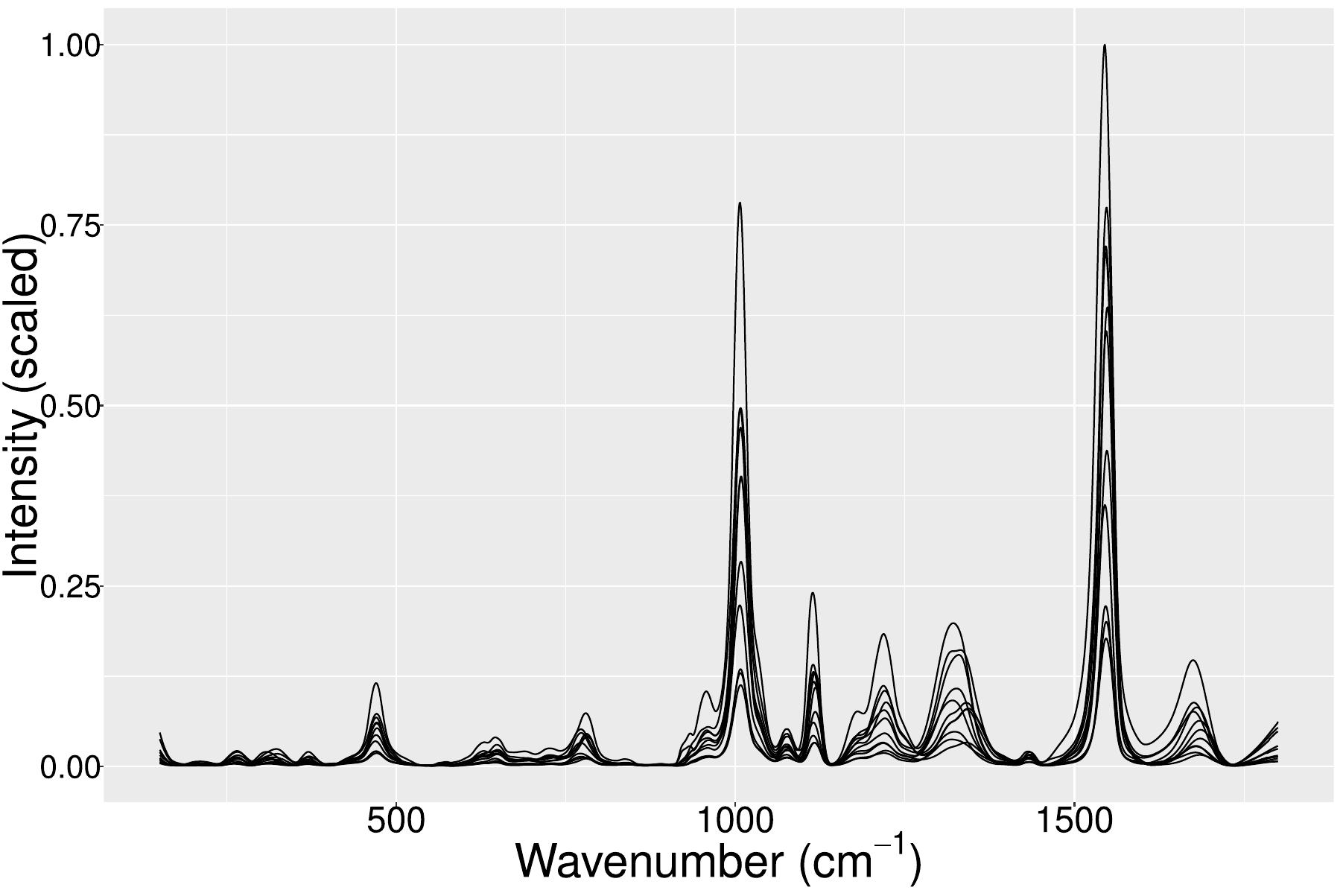}
    \caption{The mid-infrared absorbance of milk samples (left) and the surface-enhanced Raman spectra from milk samples where intensity is scaled by the maximum observed value (right).}
     \label{fig:spectral_data}
\end{figure}



\subsection{Mid-infrared spectral data for milk trait prediction} 
\label{sec:mir_milk_trait}

\begin{figure}
    \centering
    \includegraphics[width = \textwidth]{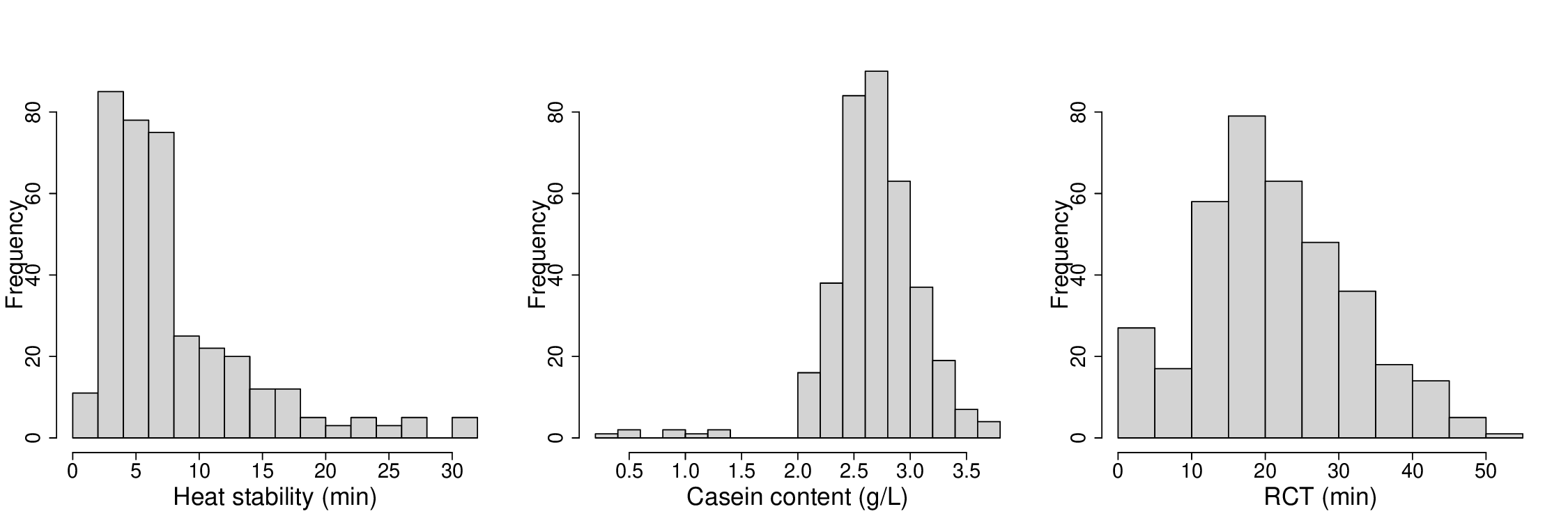}
    \caption{Histograms of the heat stability (min), casein content (gL$^{-1}$) and rennet coagulation time (RCT, min) traits from the milk MIR spectral data.}
    \label{fig:mir_traits}
\end{figure}

Mid-infrared spectral data from $730$ milk samples from $622$ individual dairy cows of multiple breeds were available; data collection protocols and trait definitions are described in detail by \cite{ViMc2015} and \cite{McVi2016}. 
The spectral data were generated from the milk samples using the MilkoScan FT6000 (Foss Electronic A/S); the instrument reports the spectrum at $1,060$ different wavenumbers. As is standard, the high-noise regions of the spectral bands relating to water ($1710$--$1600\mathrm{cm}^{-1}$, $3690$--$2990\mathrm{cm}^{-1}$ and $>3,822\mathrm{cm}^{-1}$) were omitted.  Here, interest lies in simultaneously predicting from the spectra a set of $\mathtt{R}=3$ technical and chemical traits: heat stability (min), casein content (gL$^{-1}$), and rennet coagulation time (RCT, min). In the available data, casein content is supplied by an internal black-box prediction model of the instrument, while heat stability and RCT are physically measured in a laboratory; see \cite{ViMc2015} for details. Since the prediction methods considered here assume independent and identically distributed observations, we consider only one milk sample per cow and only those milk samples for which the complete set of $\mathtt{R}=3$ traits are available. The resulting dataset, therefore, comprises $N=366$ milk samples, with $\mathtt{P}=531$ predictors and $\mathtt{R}=3$ responses. Figure \ref{fig:spectral_data} (left) shows the MIR spectra while the traits are illustrated in Figure \ref{fig:mir_traits}. This type of data is used for constructing prediction models for use by stakeholders across dairy production.

\subsection{Surface-enhanced Raman spectra from milk samples} 
\label{sec:sers_milk}

The intensity spectrum produced by surface-enhanced Raman spectroscopy (SERS) of a chemical on the surface of milk samples, along with the pH level of each sample, were collected by researchers at the Tyndall National Institute in Ireland; the resulting data are similar to those considered in \cite{WiFl2016}.

The SERS spectrum of the chemical varies with the pH of the solution in which it is immersed. The data set consists of intensity spectra of  $N=11$ milk samples at $\mathtt{P} = 1733$ wavenumbers in the range $151.422$--$1799.820\mathrm{cm}^{-1}$ and corresponding pH readings of each sample (hence here $\ttR=1$). The milk samples are from two cartons of milk, 5 and 6 samples respectively,  with best-before dates one week apart.
The cartons were kept refrigerated and spectral and pH measurements were taken daily over the course of 6 days; to reduce measurement error, each spectrum is an average of approximately $20$--$30$ measurements. Figure \ref{fig:spectral_data} (right) shows the SERS spectra; the pH values varied from $6.37$--$6.96$ with a median value of $6.87$. Interest lies in the potential of using the SERS data as a non-invasive and non-destructive means of inferring the pH of a milk sample. Such information is useful in dairy production, e.g.~to milk processors for spoilage detection or to farmers as milk pH may indicate mastitis infection \citep[e.g.][]{KaMe2019}.

\section{Model specification and inference}
\label{sec:model}
\subsection{The Bayesian partial least squares regression model} 
\label{sec:probabilistic_partial_least_squares}
Partial least squares regression assumes a linear relationship between a set of covariates and a continuous multivariate response vector through a set of latent variables whose dimension $\mathtt{Q}$ is lower than $\mathtt{P}$, the cardinality of the set of covariates. Here, we consider
PLS regression from a probabilistic modelling viewpoint and construct a principled statistical
model which assumes a similar linear relationship and imposes explicit distributional assumptions, i.e.\ we formulate the following generative model for observations $\{\bfx_n,\bfy_n\}_{n=1}^N$:
\begin{equation}
	\bfx_n = W\bfz_n+\bmeps_n, \quad \bfy_n = C\bfz_n +\bmeta_n, \quad\bfz_n\sim\normal(\OVec,\Id{\ttQ}),\label{eqn:bpls}
\end{equation}
where $W\in\Real^{\ttP\times\ttQ}$ and $C\in\Real^{\ttR\times\ttQ}$ are loading matrices, and $\bmeps_n\sim\normal(\OVec, \Sigma)$ and $\bmeta_n\sim\normal(\OVec, \Psi)$ are independent noise vectors.
Here $\normal(\boldsymbol\mu,\Phi)$ denotes a multivariate Gaussian distribution with mean vector $\boldsymbol\mu$ and a covariance matrix $\Phi$. Here, the covariance matrices $\Sigma$ and $\Psi$, are assumed to be diagonal. Using terminology from the factor analysis literature, the individual diagonal elements of $\Sigma$ and $\Psi$ are referred to as \emph{uniquenesses} and the latent $\bfz_n$ vectors are known as \emph{scores}. The specification of the latent dimension $\ttQ$ is not trivial and is discussed in detail in Section \ref{sec:inferring_the_latent_dimension}. The loading matrix $C$ can be thought of as a regression matrix; that is, in the case of $\ttR = 1$ it would be a vector of regression coefficients. As we conduct inference in the Bayesian paradigm, we refer to the model in (\ref{eqn:bpls}) as Bayesian probabilistic partial least squares (BPLS) regression and  
let $\Theta$ denote the collection of all the model parameters where $\Theta = (W,C,\Sigma,\Psi)$. The distributional assumptions in \eqref{eqn:bpls} result in the following joint density of the model variables for a single observation
\begin{align}
	\p(\bfx_n,\bfy_n,\bfz_n| \Theta) 
	&=\p(\bfx_n|\bfz_n,W,\Sigma)\p(\bfy_n|\bfz_n,C,\Psi)\p(\bfz_n)\label{eqn:lik}\\
	&\propto |\Sigma|^{-1/2} |\Psi|^{-1/2}\exp\left(-\frac{1}{2}\left(\left(\bfx_n - W\bfz_n\right)^\top\Sigma^{-1}\left(\bfx_n - W\bfz_n\right)\right.\right.\nonumber\\
	&~~~~~~~\left.\left. + \left(\bfy_n - C\bfz_n\right)^\top\Psi^{-1}\left(\bfy_n - C\bfz_n\right) + \bfz_n^\top\bfz_n\right)\right).\nonumber
\end{align}
It follows that the posterior conditional distribution for the latent variable is $\bfz_n|\bfx_n,\bfy_n,\Theta\sim\normal(\mathbf{a},S)$, where 
$S = \left(\Id{\ttQ}+W^\top\Sigma^{-1} W +  C^\top \Psi^{-1} C \right)^{-1}$ and $\mathbf{a} = S\left(W^\top\Sigma^{-1}\bfx_n +  C^\top \Psi^{-1}\bfy_n\right)$. 
Letting $\Dcal = \left( (\bfx_1,\bfy_1),\ldots,(\bfx_N,\bfy_N)\right)$ denote a dataset of $N$ independent and identically distributed observed samples,  along with latent score variables $\bfZ = (\bfz_1,...,\bfz_N)^\top$, the complete data likelihood is $\p(\Dcal,\bfZ| \Theta)  = \prod_{n=1}^N \p(\bfx_n,\bfy_n,\bfz_n| \Theta).$

For Bayesian inference, we assume a prior distribution $\pi_0$ on the model parameters and the posterior distribution follows from Bayes' Theorem: $\pi(\Theta,\bfZ|\Dcal)\propto \p(\Dcal,\bfZ| \Theta)\pi_0(\Theta)$. The choice of $\pi_0$ used here results in tractable conjugate conditional posteriors which allow exact sampling of $\Theta,\bfZ|\Dcal\sim\pi$ via a Gibbs algorithm; inference details are given in Appendix \ref{app:mcmc_details} of the Supplementary Material. The elements of $W$ and $C$ are assigned conjugate normal priors with details given in Section \ref{sec:inferring_the_latent_dimension}. The uniquenesses are assigned conjugate inverse-gamma priors: $\sigma_p^{-2}\sim\gamdist(A_\sigma,B_\sigma),~p=1,\ldots,\ttP$; and $\psi_r^{-2}\sim\gamdist(A_\psi,B_\psi),~r=1,\ldots,\ttR$.  We use $(A_\sigma,B_\sigma, A_\psi,B_\psi) = (2.5,0.1,2.5,1.5)$; the shape parameters $A_\sigma$ and $A_\psi$ are taken to be $> 1$ to ensure the first two prior moments of the uniqueness variables are finite, whereas $B_\sigma$ an $B_\psi$ are chosen to reflect the typical scale of the noise observed in the corresponding variables. Prior sensitivity analyses in typical spectral data scenarios indicated no substantial difference in results when varying $(A_\sigma,B_\sigma)$ but the uniquenesses associated with $\bfY$ required a more informative prior to avoid Heywood problems\footnote{Without an appropriately regularising prior, the Bayesian posterior of an idiosyncratic variance term (here, $\sigma^2_p,\,p=1,...,\ttP; \psi_r^2,\,r=1,...\ttR$) is multi-modal with one mode located at 0. For a detailed discussion, see Section 3.2 of \cite{FrLo2014}.}
; for all data considered here (after standardisation) our choices for $(A_\psi,B_\psi)$ were sufficiently general and stable.  An alternative approach \citep[][]{FrLo2014} is to use empirically-derived priors that vary by variable with, for example,  $B_\sigma^{p} = (A_\sigma-1)/\left[V^{-1}\right]_{pp}$, $p=1,...,\ttP$, where $V$ is the sample covariance matrix of $\bfX$; similar definitions follow for $B_\psi^r,~r=1,...\ttR$. However, experiments (not shown here) indicated that this approach generally resulted in poorer prediction accuracy. Either (or both) of the covariance matrices $\Sigma$ and $\Phi$ could be assumed to be isotropic, as is the basis for probabilistic principal component analysis \citep{TiBi1999}. This particular modification of BPLS regression (detailed in Appendix \ref{app:mcmc_details} of the Supplementary Material) gives a more parsimonious but less flexible model. 

The BPLS regression model does not make assumptions about the orthonormality of the loading matrix columns, or constrain $\ttQ$ to any specific range. This is in contrast to the probabilistic approach taken in \cite{BoUh2018} where $\ttQ\leq\min\{\ttP,\ttR\}$ is required to ensure principled frequentist inference. In BPLS regression, the need for model constraints is circumvented through the Bayesian priors which provide the required regularisation.


The BPLS regression formulation is motivated by the specific variable structure assumed in PLS. By bringing it into a probabilistic modelling paradigm, without imposing hard constraints on latent variable dimensionality or loading matrix columns, we arrive at a formulation that can be seen as a special case of a Bayesian factor analysis (BFA) model \citep[e.g.][]{PrSh1989,SoLo2001} with a blocked covariance structure. Whilst the relationship is not as direct as that established in probabilistic PCA \citep{TiBi1999}, it nonetheless admits the emulation of PLS using a probabilistic model. We exploit this connection by implementing a Bayesian nonparametric prior commonly seen in BFA models to avoid the need to specify the latent dimension (see Section \ref{sec:inferring_the_latent_dimension}).
Additionally, 
by drawing on the regression aspect of (\ref{eqn:bpls}), we can appropriately modify parts of the model using sparsity-inducing priors to improve overall predictive performance (see Sections \ref{sec:spike_and_slab_regression} and \ref{sec:bayesian_lasso_regression}).
While \cite{Viva2013} also proposes a Bayesian PLS-type model, the assumed data-generating mechanism does not emulate the relationship between the predictors and the response in the same way as standard PLS regression, and so is fundamentally different to the BPLS regression model proposed here. 
Furthermore, \cite{Viva2013} carry out approximate variational inference, with no quantification of the mean field approximation error in the predictions provided.

Regarding identifiability, an inherent feature of factor models is that the loading matrices, $W$ and $C$, and scores $\bfZ$ are not identifiable---rotating these in the latent space results in the same posterior density. However, while Procrustean rotations \citep{HoffEtAl2002, McParlandEtAl2017} could be employed to resolve this, the quantities of interest here are the posterior predictive distributions of the response vectors which are invariant to rotations (see Section \ref{sec:predictive_distribution}) and so such an approach is possible but unnecessary.

\subsection{Inference on infinite latent dimensions} 
\label{sec:inferring_the_latent_dimension}
Methods based on dimension-reduction such as PLS generally require the user to select the dimension of the latent variables (often termed the number of components) in the model; the choice can often be subjective, or based on data-sensitive heuristics. A common approach for PLS is to use cross-validation (CV) in order to minimise the root mean squared error (RMSE) of the predictions; this is the default in standard statistical packages such as \texttt{pls} \citep{LiMe2022} and \texttt{spls} \citep{ChCh2019} in R \citep{R2023}. These methods require many separate model fits and one needs to specify the data folds which involves a trade-off between metric robustness and computational efficiency.  Furthermore, there is no clear consensus as to how to choose $\ttQ$ when fitting a given model to multivariate responses, as is required in the motivating applications here. In practice, one either needs to fit each response variable separately or use subjective judgement when choosing a joint model. 

In the proposed BPLS regression framework, we avoid the selection process and fit only one model by using an appropriate nonparametric prior. This obviates the need to choose $\ttQ$ and allows for posterior predictions which reflect its inherent uncertainty, thus making the framework more objective, setting-agnostic and user-friendly. We employ the popular multiplicative gamma process (MGP) prior \citep{BhDu2011} which assumes the effective dimensionality $\ttQ$ is infinite whilst enforcing shrinkage on the entries of the loading matrices ($W$ and $C$) which increases with the column index. This is achieved by taking the prior precisions of the loadings to be a stochastically increasing sequence of gamma variables. Specifically, the MGP prior considered here is:
\begin{align*}
	w_{pq}|\phi_{pq},\tau_q&\sim\normal(0,\phi_{pq}^{-1}\tau_q^{-1}),~ \phi_{pq}\sim\gamdist(\nu^w_1,\nu^w_2),~~p=1,...,\ttP,~q\in\mathbb{N};\\
	c_{rq}|\dot\phi_{rq},\tau_q&\sim\normal(0,{\dot\phi_{rq}}^{-1}{\tau_q}^{-1}),~\dot\phi_{rq}\sim\gamdist(\nu^c_1,\nu^c_2),~~r=1,...,\ttR,~q\in\mathbb{N}; \mbox{ where}\\
\tau_{q}&=\prod_{k=1}^{q}\delta_k,\quad
\delta_1=1,~\mbox{and}~\delta_k\sim\gamdist(\alpha,\beta),~k=2,...,q.
\end{align*}
The conditional posteriors of the hyperparameters $(\boldsymbol\phi,\dot{\boldsymbol\phi},\boldsymbol\delta)$ are tractable and can be inferred through a Gibbs sampling algorithm. The prior formulation considered here differs from that of \cite{BhDu2011} in that, without loss of generality, we fix $\delta_1$ as all prior precisions are multiplied by the term---empirically no significant effect on the final model fit was observed. The MGP prior has been widely used in latent variable models \citep[e.g.][]{OvAb2016,MoVi2020}. The $\alpha$ parameter could be inferred; however, we observed that this can lead to unstable results. Whilst $\alpha>\beta$ leads to stochastically increasing $\bmtau$, the reciprocals of its elements are not guaranteed to be stochastically decreasing. To overcome the issue, \cite{Dura2017} advises fixing $\beta=1$ and $\alpha>\beta+1$ for reliable analyses. There is more flexibility when specifying the individual precision components $\phi_{pq}$ and $\dot\phi_{rq}$; adopting hyperparameters such that $\mathbb{E}[\phi_{pq}^{-1}],~\mathbb{E}[\dot\phi_{rq}^{-1}]\leq1$ allows for sparser loading matrices, implying that $\nu^w_2\leq \nu^w_1-1$ and $\nu^c_2\leq \nu^c_1-1$. Here, we set $(\alpha,\beta,\nu^w_1,\nu^w_2,\nu^c_1,\nu^c_2) = (2.2,1,2,3,2,3)$; prior sensitivity analyses showed no substantial difference in results when varying these, provided the recommendations are followed. 

For practical implementation of the MGP prior, we consider truncation at some appropriate dimension $\ttQ^*$; \cite{BhDu2011} and \cite{GwGo2024} suggest using a reasonably large value. Here, at the initialisation stage of inference, standard principal component analysis is performed on $\bfX$ and a conservative bound $\ttQ^*$ on the effective dimension is obtained; e.g.\ one which explains $0.99$ of the observed variation.  For the spectral data considered here, this procedure gave $\ttQ^*$ between 8--27. This selection can be verified by examining the columns of the loading matrices  during warm-up runs of the Markov chain Monte Carlo (MCMC) chains; the magnitudes of the elements of the latter columns should be, for example, less than 1\% of the first few columns indicating negligible contributions to the variation in the data. For the motivating datasets here, we found that 5,000 iterations (after discarding the burn-in) were sufficient for reliably diagnosing the choice of $\ttQ^*$. Additionally, the posterior distribution of $\bmtau$ can be used to determine if the shrinkage prior hyperparameters are appropriate; see Appendix \ref{app:posterior_shrinkage} of the Supplementary Material. 
In the numerical experiments in Section \ref{sec:numerical_experiments}, we consider synthetic data generated using $\ttQ=10$ latent variables and for inference we truncate the infinite model at $\ttQ^*=15$. This is relatively large compared to what appears in, e.g.\  \cite{BoUh2022} and is motivated by the intricacy of spectral data; dimension-reduction methods applied to spectral data in Sections \ref{sec:benchmark} and \ref{sec:data_examples} generally required between 5 and 20 latent variables. 

An adaptive Gibbs algorithm such as that in \cite{BhDu2011} could be used to automate the process of choosing $\ttQ^*$, however, here preliminary experiments showed no substantial computational gain from using the adaptive scheme as the adaptive steps increased the resulting MCMC burn-in. 

\subsection{Spike-and-slab BPLS regression} 
\label{sec:spike_and_slab_regression}

It is possible that some of the leading principal components of $\bfX$, accounting for much of its variation, play no role in the relevant regression in \eqref{eqn:bpls}. Indeed, this is the basis for orthogonal partial least squares approaches \citep[O2-PLS][]{TrWo2002,TrWo2003}, where the latent variables are split into \emph{joint} components, shared between both $\bfX$ and $\bfY$, and \emph{specific} components, split into two further sets: ones only interacting with $\bfX$ and ones only interacting with $\bfY$. A probabilistic extension PO2-PLS \citep{BoUh2022} has a necessary constraint $\ttR\leq\ttQ$ and, notably, this further requires the user to choose the dimensionality of each of the component sets. 

The decision of whether a given component is shared between the two observed variable sets or only explains one equates to a variable selection problem on the space of the latent variables. We therefore propose an extension to BPLS which emulates O2-PLS by employing a  modification based on spike-and-slab regression \citep{EdRo1993}. Here, as we strictly predict $\bfy$ given $\bfx$, we only consider specific components relating to $\bfx$ which corresponds to some columns of regression matrix $C$ being set to zero. We augment the model with a binary, diagonal matrix $B=\mathsf{diag}(b_1,\ldots,b_\ttQ)$, whose diagonal entries are independent Bernoulli random variables, each with a prior probability of success $p_0\sim\unif(0,1)$. The binary variates act as ``switches'' for their respective variables. The matrix is incorporated into the relevant regression part of \eqref{eqn:bpls}, i.e.\ $\bfy = CB\bfz +\bmeta$, with the product $CB$ providing the desired sparsity. The posterior sampling of $\{b_q\}_{q=1}^\ttQ$ is based on exploiting the product form of the Gaussian likelihood \eqref{eqn:lik} as detailed in Appendix \ref{app:gibbs_ss_bpls} of the Supplementary Material. We call this model variant ss-BPLS.  

While the orthogonal PLS methods do not necessarily improve upon the predictions of standard PLS \citep{TrWo2002} they can, however, offer more interpretable results by discarding parts of $\bfx$ which are redundant in explaining the variation in $\bfy$. However, due to the rotational invariance of a Gaussian factor model, there is a lack of identifiability in the $B$ matrix in the absence of hard constraints. Hence, whilst the posterior distribution of $B$ here may not be useful in analysing the relationship between $\bfX$ and $\bfY$, the spike-and-slab mechanism can assist in avoiding overfitting where too many latent variables are required to explain the variation in the response vector. Sparse regression can aid the estimation of the $\Psi$ matrix elements, producing more reliable prediction intervals.

\subsection{LASSO BPLS regression} 
\label{sec:bayesian_lasso_regression}

As we are interested in modelling multivariate $\bfy$ vector, some latent variables may affect one subset of the responses and not another, implying sparsity on the individual elements of $C$, not on entire columns of $C$. This could be achieved by a LASSO-type penalty on the values of the entries \citep[e.g.][]{ChOe2010}. An explicit Bayesian prior of the form $\pi_0(C)\propto \exp\left(-\lambda\sum_{r,q} |c_{rq}|\right)$, where $\lambda>0$ is a regularisation hyperparameter, leads to an intractable posterior density severely complicating the inference. However, as outlined in \cite{PaCa2008}, setting up a Bayesian prior structure  $	c_{rq}\sim\normal(0,\dot\phi_{rq}^{-1}\tau_q^{-1})~\mbox{and}~\dot\phi_{rq}^{-1}\sim\Exp \left(\frac{\lambda^2}{2}\right),$ can give the desired $\ell_1$-penalty. It can be shown that the marginal prior density of $c_{rq}$ has the form	$\pi_0(c_{rq}|\lambda,\boldsymbol{\tau})\propto\frac{\lambda}{2\sqrt{\tau_q^{-1}}}\mathsf{e}^{-\lambda|c_{rq}|/\sqrt{\tau_q^{-1}}},$ which leads to tractable conditional posteriors; \emph{a posteriori}, the individual $\dot\phi_{rq}$ parameters are inverse-Gaussian distributed. Furthermore, assigning a prior distribution $\lambda^2\sim\gamdist(A_\lambda,B_\lambda)$ leads to a conjugate conditional posterior, thus allowing inference of the regularisation hyperparameter; here, we use $A_\lambda=B_\lambda=1$, but  other choices could be appropriate for specific settings. The details of the resulting MCMC scheme are given in Appendix \ref{app:gibbs_l_bpls} of the Supplementary Material. We refer to this formulation of the model as L-BPLS. 

\cite{Rock2023} argues that point predictions produced from the marginal posterior predictive of a Bayesian LASSO model have an inherent bias and are prone to overfitting as the number of variables increases. However, the same work shows that if the hierarchical LASSO-type prior is combined with another sparsity prior (e.g.~spike-and-slab mixture) then intended predictive performance is recovered. Here, we employ the MGP shrinkage prior which enforces shrinkage on the latter columns of the loading matrices and, as seen in the simulation studies in Sections \ref{sec:numerical_experiments} and \ref{sec:data_examples}, accurate predictive performance is indeed observed.

\subsection{Predictive distribution of the response} 
\label{sec:predictive_distribution}
The aim of BPLS regression is to accurately characterise the posterior predictive distribution of response traits $\bfy_+$ given only new spectral samples $\bfx_+$; such distributions are readily accessible through the Bayesian framework. We first note the conditional distributions involving the latent variable $\bfz_+$, i.e.  $\bfz_+|\bfx_+,\Theta\sim\normal(S_zW^\top\Sigma^{-1}\bfx_+,S_z)$ where $ S_z = \left(\Id{\ttQ}+ W^\top\Sigma^{-1} W\right)^{-1}$ which follows from the incomplete likelihood where only $\bfx_+$ is observed and $\bfy_+|\bfz_+,\Theta\sim\normal(C\bfz_+,\Psi)$ which follows from the assumed generative model \eqref{eqn:bpls}. Integrating out the latent $\bfz_+$, we find the marginal posterior predictive distribution  $\p(\bfy_+|\bfx_+,\Theta)$ is also Gaussian with  $\mathbb{E}[\bfy_+|\bfx_+,\Theta] = CS_zW^\top\Sigma^{-1}\bfx_+ $ and $\mathbb{V}[\bfy_+|\bfx_+,\Theta] = CS_zC^\top+\Psi$. Working in the Bayesian paradigm allows us to incorporate the uncertainty in $\Theta$ into the predictions by marginalising over the parameter posterior, $\p(\bfy_+|\bfx_+,\Dcal) = \int \p(\bfy_+|\bfx_+,\Theta)\pi(\Theta|\Dcal)\;\mathsf{d}\Theta$.  Expectations with respect to these marginal posterior predictive distributions can be approximated via Monte Carlo integration based on the MCMC output; Appendix \ref{app:estimating_the_moments_of_the_posterior_predictive} of the Supplementary Material details this procedure. 

In what follows, the point predictions from BPLS regression models are given as $\mathbb{E}[\bfy_+|\bfx_+, \Dcal]$ but other options are available such as predicting from the posterior mode; a comparison in Appendix \ref{app:posterior_mode} of the Supplementary Material shows there is little substantive difference between the approaches. The prediction intervals for each observation are constructed from equal-tail quantile intervals of the posterior predictive distribution. These are very similar to the highest posterior density intervals; the predictive distributions are unimodal and symmetric due to the distributional assumptions of the model.

\section{Numerical experiments} 
\label{sec:numerical_experiments}


In this section, we evaluate the empirical performance of the BPLS regression framework on synthetic and benchmark data and compare its performance to other well-utilised methods: partial least squares where $\ttQ$ is chosen through cross-validation and the ``one-sigma'' heuristic approach \citep{Voet1994}, denoted PLS and PLS-1s respectively; sparse partial least squares (sPLS); principal component regression (PCR); and LASSO \citep{Tibs1996} and ridge regressions \citep{HoKe1970}. The number of components for PLS and PCR methods was chosen using the default 10-fold cross-validation using the \texttt{pls} package in \textsc{R}; sPLS used the default cross-validation in the \texttt{spls} package. We assess performance through the root mean squared error of the predictions (RMSEP) on a given test set. The probabilistic partial least squares (PPLS) regression model \citep{BoUh2018} is not applied to the synthetic data here as a setting where $\ttR<\ttQ$ is of interest, which PPLS does not permit. The PPLS model is however used in comparisons on real data where the latent dimension is unknown (see Sections \ref{sec:benchmark} and \ref{sec:data_examples}).

The Gibbs sampling algorithm is used for BPLS regression inference, with initial values and the iterations of the Markov chains detailed in Appendix \ref{app:mcmc_initialisation} of the Supplementary Material. For synthetic data examples, a conservative choice of $\ttQ^*=15$ was used when inferring BPLS models; for analyses on real spectral datasets we employed the method discussed in Section \ref{sec:inferring_the_latent_dimension}. To assess convergence, we consider the mixing of the entries of $\widehat{\bfY}^{\mathrm{test}} =  CS_zW^\top\Sigma^{-1} \bfX^{\mathrm{test}}$; this product is invariant to the different rotations of the latent space. The chains are run long enough such that after appropriately discarding the burn-in, we are left with at least $1,000$ effective samples of each element. This condition indicates the chains are mixing sufficiently well; see, e.g.\ \cite{DoKn2021}. Here, this was safely achieved in all scenarios with a burn-in of $9,000$ iterations and a further $21,000$ iterations. The BPLS models are fitted to multivariate responses whereas the competitors are fitted to each univariate response element; preliminary experiments indicated no benefit in using multiple univariate-response BPLS regression models.

\subsection{Synthetic datasets}
\label{sec:synthetic}

The performance of BPLS regression is assessed by comparison to that of competitors across different data scenarios. To emulate the motivating practical applications, we consider a multivariate response where $\ttR = 4$ and set $\ttQ=10$ with $N \in\{50,500\}$ and $\ttP\in\{100,1000\}$.  The loading matrices' entries are simulated using via $w_{pq}\stackrel{\mathrm{iid}}{\sim}\normal(0,1/\ttQ)$ and $c_{pq}\stackrel{\mathrm{iid}}{\sim}\normal(0,1/\ttQ)$; this controls the marginal prior variance contribution to the components of $\bfx$ and $\bfy$. We consider two isotropic noise scenarios: low-noise and high-noise, corresponding to $\sigma^2=\psi^2=0.1$ and $\sigma^2=\psi^2=0.5$ respectively. Appendix \ref{app:sim_sparse} of the Supplementary Material provides additional results where the $C$ matrix is sparse; all BPLS regression variants were in fact robust in such settings.

Under the hyperparameter settings outlined in Section \ref{sec:inferring_the_latent_dimension}, each model is fitted to a training dataset of size $N_{\mathrm{train}}$, simulated with the above specifications and then standardised. The fitted models are then used to make predictions for $N_{\mathrm{test}} = 1,000$ unseen test samples, and the RMSEP is given as the average error over the response elements; by construction, they are on the same scale. This setup is repeated for a total of $10$ synthetic data sets. 

\begin{figure}[tb]
	\centering
	\includegraphics[width = \textwidth]{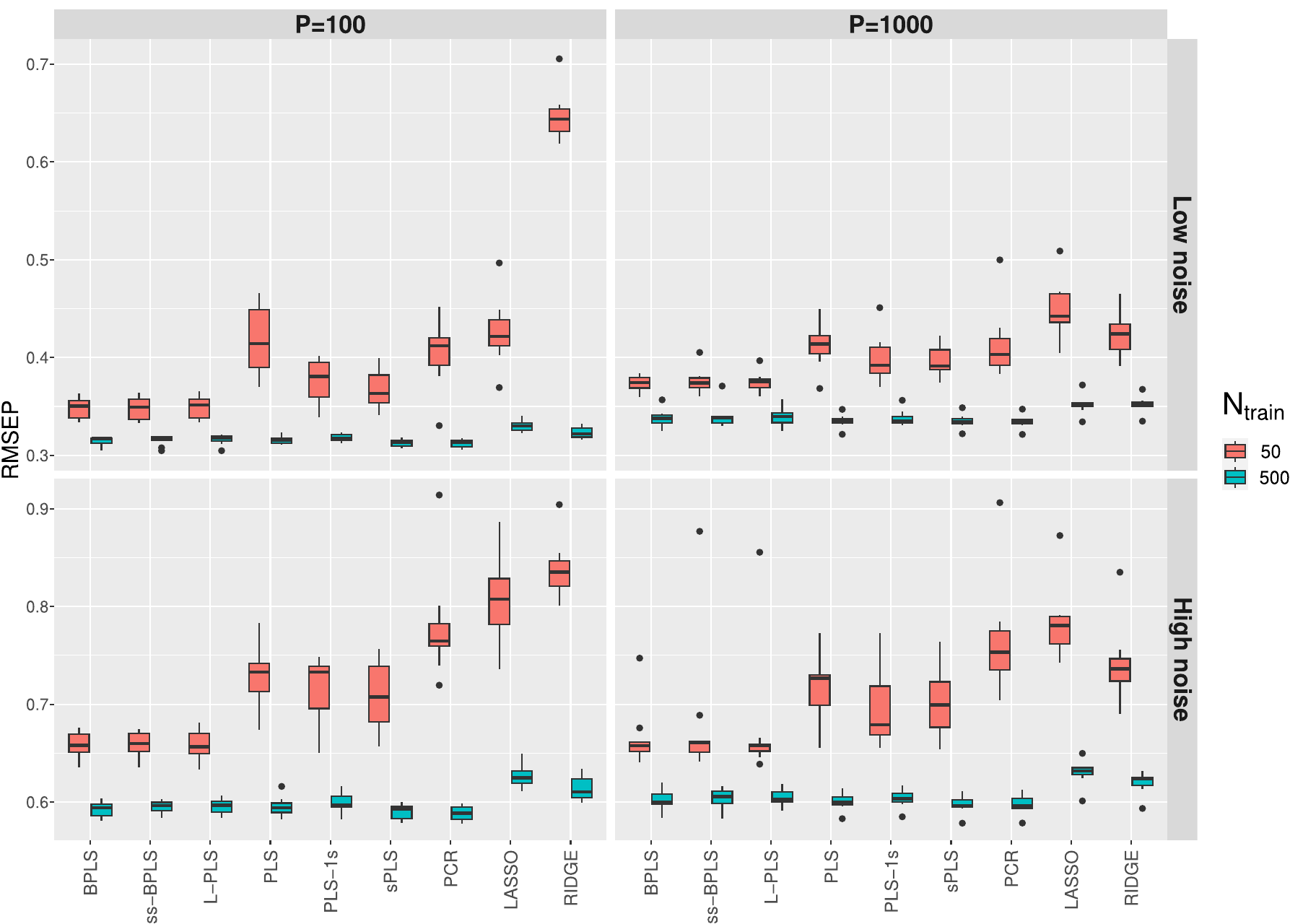}
	\caption{The RMSEP across 10 synthetic datasets for different settings of $N_{\mbox{train}}$ and $\mathtt{P}$ under the BPLS and existing methods. Low-noise and high-noise scenarios correspond to $\sigma^2=\psi^2=0.1$ and $\sigma^2=\psi^2=0.5$ respectively.}
	\label{fig:simboxplotfree}
\end{figure}

Figure \ref{fig:simboxplotfree} illustrates how the BPLS performance compares to other methods. In scenarios where the data have small $N_{\mathrm{train}} = 50$ and $\ttP = 100$, the BPLS, ss-BPLS and L-BPLS models provide a clear advantage in terms of prediction accuracy over their competitors. As the number of training samples increases to $N_{\mathrm{train}} = 500$, all PLS-type methods perform similarly and tend to be better than the standard LASSO and ridge regression competitors. Figure \ref{fig:simboxplotfree} shows that the impact of the nonparametric prior resulting in bias in an infinite-dimensional model does not necessarily disappear as the sample size grows \citep{DiFr1986,Rous2016}. The presence of high levels of noise decreases the accuracy of the competing methods more substantially as compared to the BPLS regression approaches.  Appendix \ref{app:posterior_shrinkage} of the Supplementary Material provides details on the shrinkage parameter posteriors showing the BPLS regression model correctly identifies the latent dimensionality of the observed data. An additional comparison of the BPLS methods to standard Bayesian factor analysis on the most challenging synthetic data setting (i.e.,\ $\ttP=1000,\,N=50,\,$high noise) is given in Appendix F.1 of the Supplementary Material.

In addition to strong predictive performance, the BPLS models provide uncertainty quantification through the posterior predictive distributions. Table \ref{tab:sim_coverage} gives the estimated coverage probabilities for test-set observations under various simulation settings. All models achieve the desired coverage when larger samples are available ($N_{\mathrm{train}}=500$), but when only a small number of samples is available, the prediction interval (PI) width is under-estimated for BPLS and ss-BPLS.  Upon further investigation, the too narrow PIs correspond to under-estimation of the $\Psi$ matrix. The PIs produced by L-BPLS are a little too conservative when the noise is low but too narrow when it is high; indeed, this suggests that a sparse $C$ matrix leads to a $\Psi$ posterior which relies heavily on the prior. There did not appear any meaningful difference in the prediction interval widths achieving the same coverage. 

\begin{table}[]
\caption{Estimated prediction coverage of 95\% prediction intervals produced by different BPLS regression models under different data settings; \textsf{mean (sd)}. At each iteration the median coverage value is taken across the elements of the multivariate response vector. Low-noise and high-noise scenarios correspond to $\sigma^2=\psi^2=0.1$ and $\sigma^2=\psi^2=0.5$ respectively.}
\begin{tabular}{rl|ll|ll}
        &       & \multicolumn{2}{c|}{\textbf{Low noise}} & \multicolumn{2}{c}{\textbf{High noise}} \\
        &       & $\ttP=100$         & $\ttP=1000$         & $\ttP=100$          & $\ttP=1000$         \\ \hline
\textbf{BPLS}    & $N=50$  & $0.929~(0.014)$&$0.880~(0.026)$&$0.911~(0.029)$&$0.844~(0.056)$\\
        & $N=500$ & $0.948~(0.007)$&$0.945~(0.005)$&$0.945~(0.007)$&$0.945~(0.008)$\\ \hline
\textbf{ss-BPLS} & $N=50$  & $0.927~(0.014)$&$0.891~(0.029)$&$0.906~(0.028)$&$0.845~(0.041)$\\
        & $N=500$ & $0.948~(0.009)$&$0.945~(0.004)$&$0.945~(0.006)$&$0.944~(0.007)$ \\ \hline
\textbf{L-BPLS}  & $N=50$  & $0.966~(0.008)$&$0.962~(0.009)$&$0.938~(0.021)$&$0.937~(0.019)$\\
        & $N=500$ & $0.952~(0.005)$&$0.952~(0.002)$&$0.950~(0.005)$&$0.951~(0.002)$\\ 
\end{tabular}\label{tab:sim_coverage}

\end{table}

\subsection{Benchmark spectral data}
\label{sec:benchmark}
Here, we compare existing approaches and the proposed BPLS regression methods on near-infrared (NIR) spectral data of grain mash samples considered in \cite{LiFr2009}; data are available from the \textsc{R} package \texttt{chemometrics} \citep{FiVa2017}. Each of the $N=166$ samples in the data contains $\mathtt{P}=235$ predictor variables, given as the approximate first derivative of the curve made up of absorbance values, and $\mathtt{R}=2$ traits: ethanol concentration and glucose concentration (both  measured in gL$^{-1}$). This data set is well utilised as a benchmark for dimension-reduction prediction methods \citep[e.g.][]{FiGs2012}. 

We consider a similar setup to that in \cite{FrGo2021}: the dataset is divided into four approximately  equal folds, and three folds are used as the training set $\Dcal^{\mathrm{train}} = \left(\bfX^{\mathrm{train}},\bfY^{\mathrm{train}}\right)$ with $N_{\mathrm{train}}$ observations and the remaining fold is treated as the test set, $\Dcal^{\mathrm{test}} = \left(\bfX^{\mathrm{test}},\bfY^{\mathrm{test}}\right)$ with $N_{\mathrm{test}}$ observations; all four fold combinations are considered.  As in Section \ref{sec:synthetic}, the BPLS regression models are fitted to multivariate response vectors, whereas  competitor methods each involve multiple univariate response fits. In addition to the competitor methods used in Section \ref{sec:synthetic}, we also consider the frequentist probabilistic partial least squares model \citep[PPLS,][]{BoUh2018} fitted to the multivariate response. In PCR, we identify the components with significant regression coefficients, and in BPLS regression models, we let $\widehat{\ttQ}_{y_r}$ be the effective dimension of the regression of response element $y_r$ ($r=1,\ldots,\ttR$), i.e.~$\widehat{\ttQ}_{y_r} = \sum_{q=1}^{\infty} \mathbb{I}_{\{|\hat{c}_{rq}|\geq 0.05c^*\}},$ where $\hat{c}_{rq}$ are the posterior mean estimates of the entries of $C$ and $c^* = \max_{r,q}|\hat{c}_{rq}|$. Performance metrics are given as empirical averages of the RMSEP across the four fold combinations, and we also report the estimates of the effective dimensionality of the data (i.e. number of latent variables or components) for each method as an average over the four folds. 

As shown in Table \ref{tab:nir_grain}, in the benchmark grain mash NIR spectral data, all three of the BPLS-based methods outperform the existing standard approaches in terms of prediction accuracy, with the L-BPLS approach performing optimally. The BPLS methods had lower variation in the $\widehat{\ttQ}_{y_r}$ estimates which suggests that the inaccuracy for other methods here may be due to the incorrect choice of the latent dimension. A comparison of BPLS methods to standard Bayesian factor analysis approaches on the NIR dataset can be found in Appendix F.1 of the Supplementary Material.  Additionally, Figure \ref{fig:NIRgrain_PredIntervals} illustrates point predictions and 95\% prediction intervals from the L-BPLS model for 40 random test samples in one test fold of the NIR grain spectral data. The natural provision of coherent prediction intervals under the BPLS approach, particularly in the case of glucose (Figure \ref{fig:NIRgrain_PredIntervals} (left)), provides richer inference, enabling more informed decision-making. For example, as noted in \cite{LiFr2009}, it is desirable to dilute the grain mash to achieve a glucose concentration of $\leq$1gL$^{-1}$ for fermentation. If NIR spectral measurements are used instead of more expensive gold standard tests, there will be  uncertainty in the final glucose predictions. Treating these predictions as infallible could lead to grain mash samples which have too high a concentration of glucose for processing purposes. To mitigate this, the processor could instead employ a more conservative dilution protocol using an upper quantile of the posterior predictive  distribution as the glucose content prediction, thereby ensuring robustness.

\begin{figure}[tb]
	\centering
     \includegraphics[width = 0.49\textwidth]{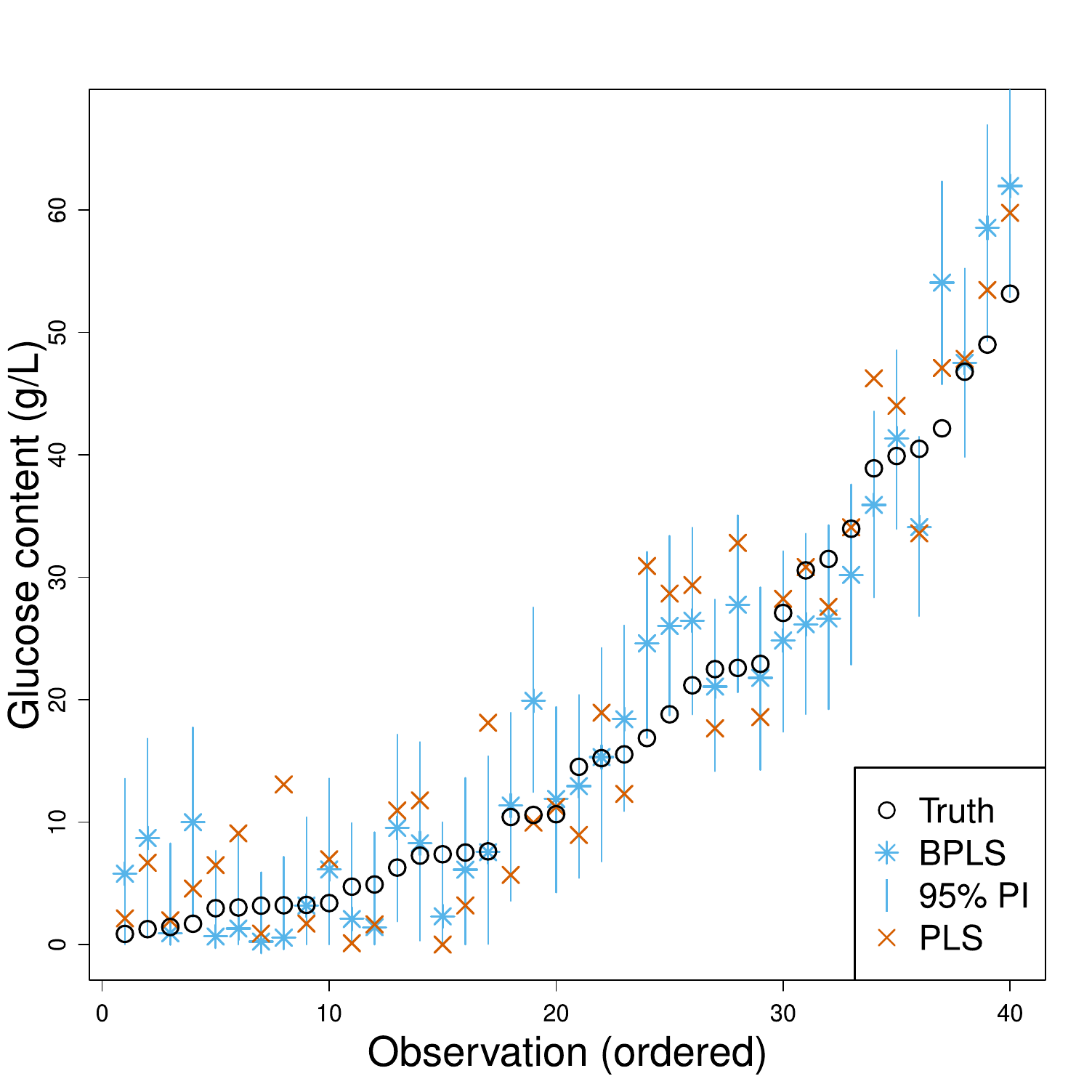}
     \includegraphics[width = 0.49\textwidth]{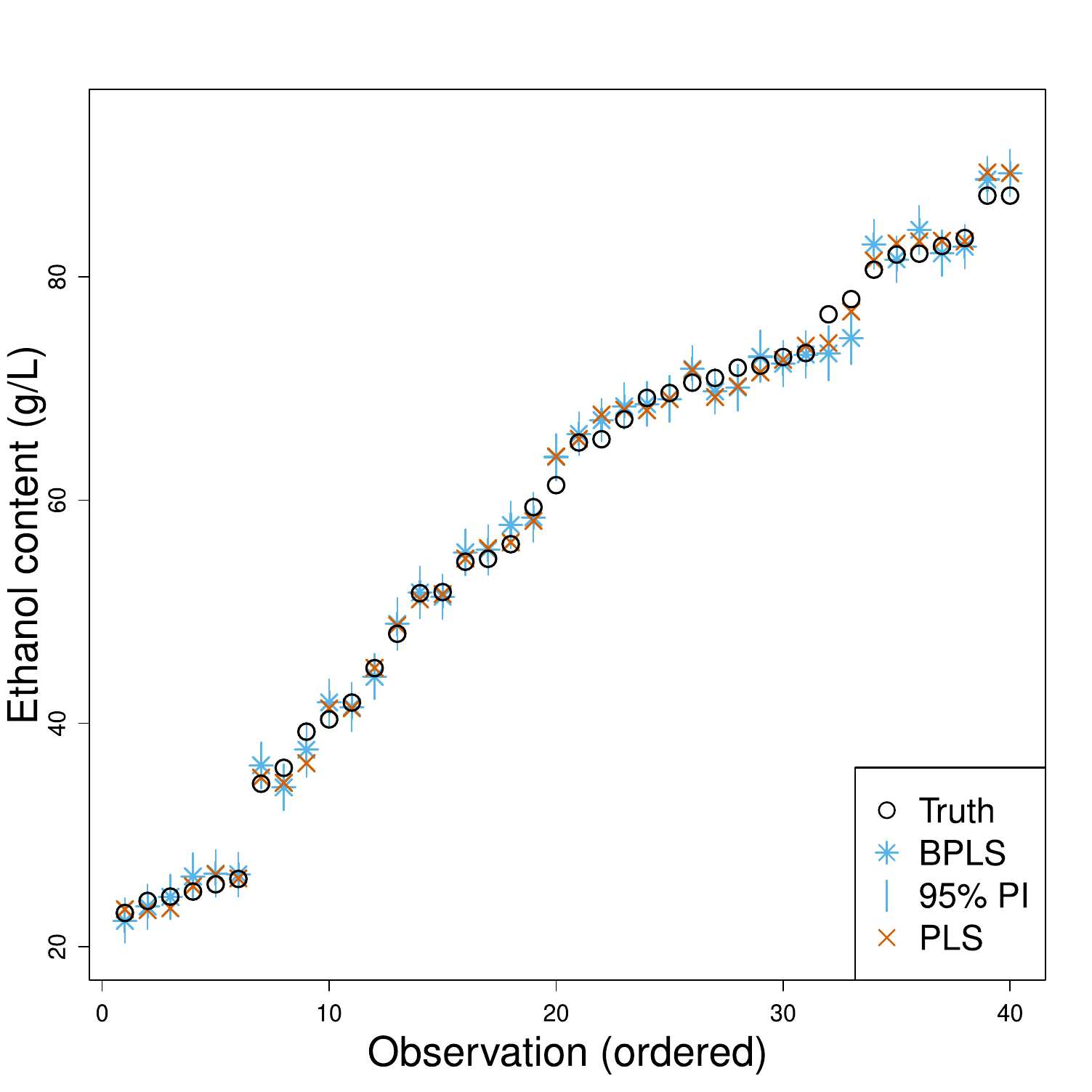}
	\caption{Point predictions and $95\%$ prediction intervals of glucose and ethanol content from the L-BPLS model in one test fold of the NIR grain spectra dataset; 40 random test samples are given in increasing order of respective trait quantities for clarity.  In addition, true values and predictions from standard PLS are provided.}
	\label{fig:NIRgrain_PredIntervals}
\end{figure}

\begin{table}[tb]
	\caption{Accuracy of predictions of the standardised glucose and ethanol traits under the proposed BPLS and existing methods on the benchmark grain mash NIR spectral data. Estimates are given as `mean (standard deviation)' over the four test folds, with bold font indicating optimal performance. $^*$Due to numerical instability, $\widehat{\ttQ}_y$ was picked manually.}
	\label{tab:nir_grain}
	\centering

	\begin{tabular}{l|cc|cc}
	\hline

	\hline
	&\multicolumn{2}{c}{\textbf{Glucose}} & \multicolumn{2}{|c}{\textbf{Ethanol}} \\
	&RMSEP&$\widehat{\ttQ}_y$& RMSEP&$\widehat{\ttQ}_y$  \\
	\hline
BPLS & 0.358 (0.045) & 8.5 (0.6) & 0.061 (0.009) & 7.5 (1.3) \\ 
  ss-BPLS & 0.395 (0.041) & 9.0 (1.4) & 0.061 (0.005) & 7.3 (0.5) \\ 
  L-BPLS & \textbf{0.353} (0.053) & 14.0 (0.8) & \textbf{0.060} (0.008) & 8.3 (0.5) \\   \hline
  PLS & 0.422 (0.048) & 8.3 (1.3) & 0.072 (0.012) & 10.3 (1.3) \\ 
  PLS-1s & 0.447 (0.043) & 10.8 (3.6) & 0.066 (0.009) & 10.5 (1.7) \\ 
  sPLS & 0.442 (0.065) & 16.0 (3.7) & 0.070 (0.008) & 13.0 (2.3) \\ 
  PCR & 0.440 (0.038) & 18.0 (1.2) & 0.069 (0.006) & 22.3 (2.5) \\ 
  LASSO & 0.428 (0.018) & -- & 0.080 (0.012) & -- \\ 
  RIDGE & 0.727 (0.033) & -- & 0.264 (0.031) & -- \\ 
		PPLS & 0.923 (0.064) & $^*$1.0 (0.0) & 0.491 (0.038) & $^*$1.0 (0.0)\\
	\hline

	\hline
	\end{tabular}
\end{table}

\section{Predicting milk traits from MIR and SERS spectral data} 
\label{sec:data_examples}

We use the developed BPLS regression methods to predict milk traits from the motivating spectral datasets described in Section \ref{sec:data_descriptions}. We use the same hyperparameter settings and MCMC diagnostics as in Section \ref{sec:numerical_experiments} and we use the same four-fold training-test approach and compare the different methods in the same way as detailed in Section \ref{sec:benchmark}.

\subsection{Predicting technological and protein traits from MIR milk spectral data} 
\label{sec:data_examples_MIR}

\begin{figure}[tb]
	\centering
	\includegraphics[width = 0.32\textwidth]{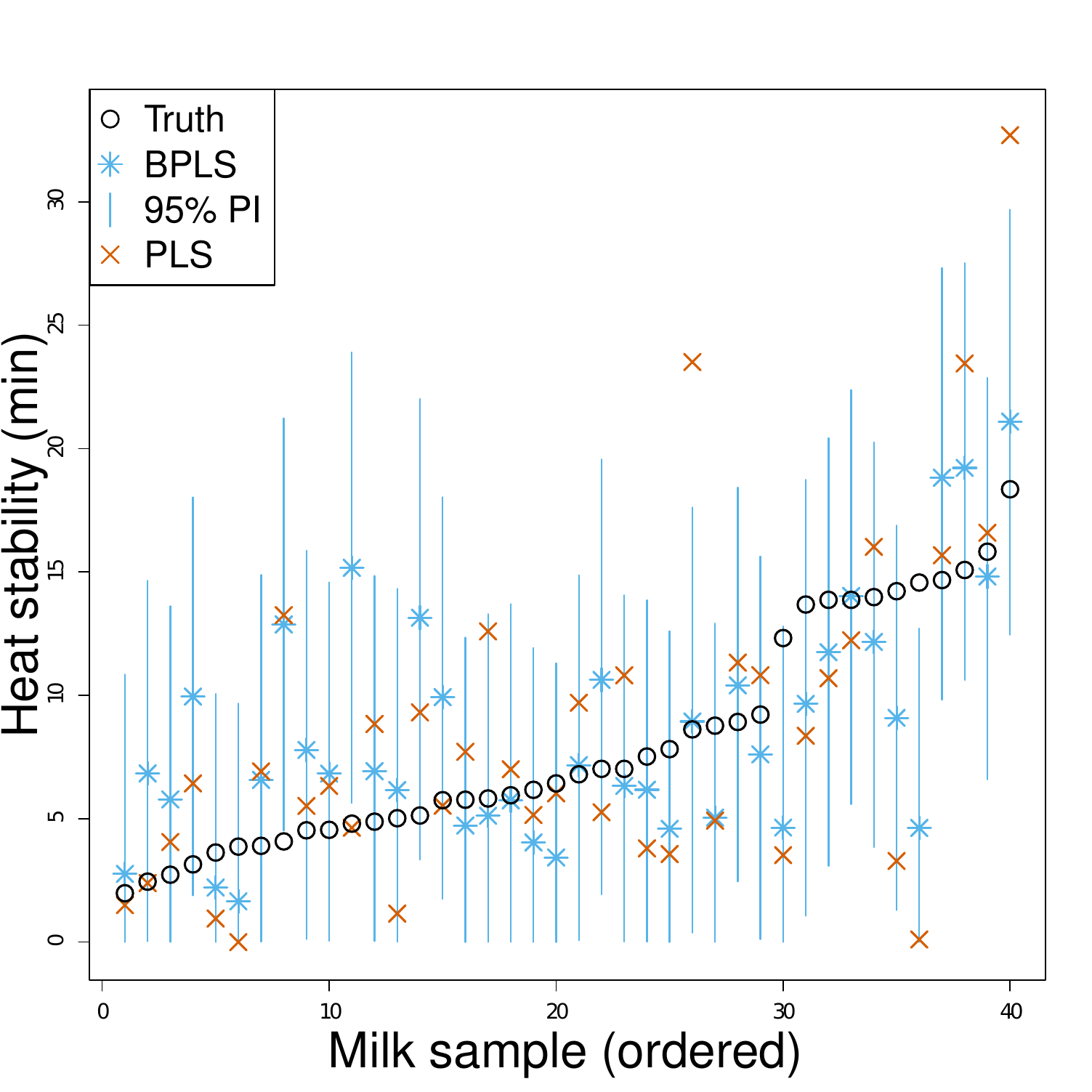}
 \includegraphics[width = 0.32\textwidth]{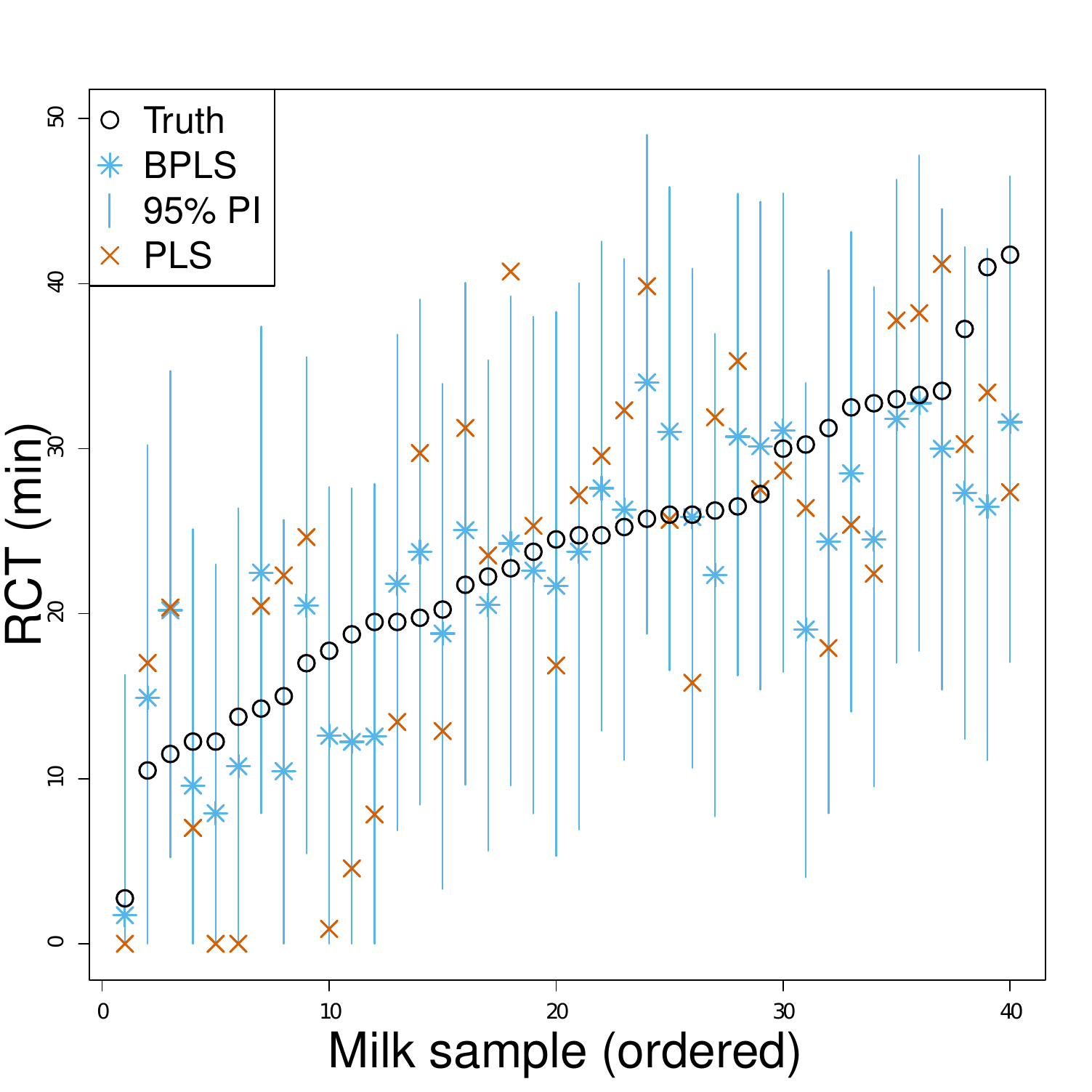}
  \includegraphics[width = 0.32\textwidth]{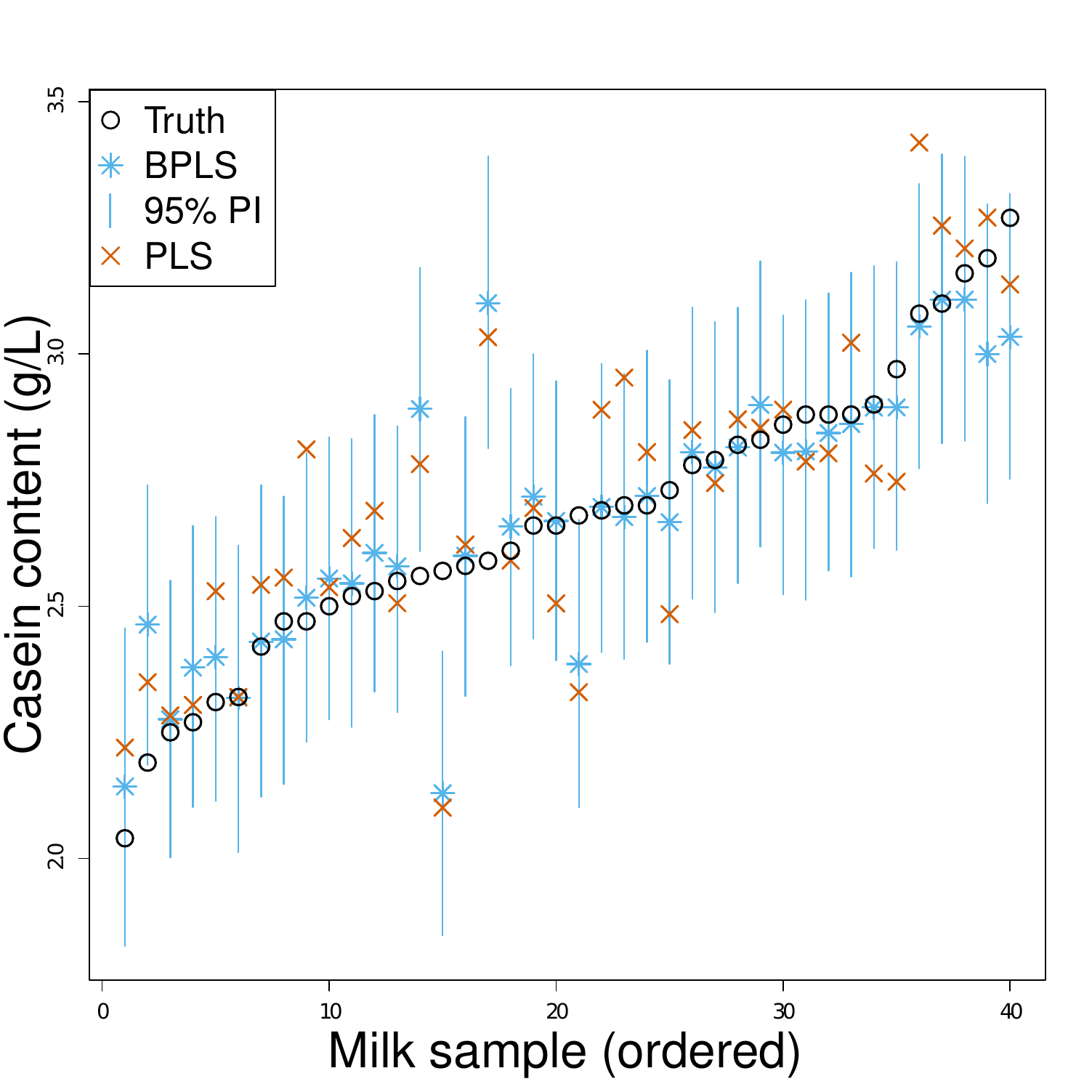}
	\caption{Point predictions and $95\%$ prediction intervals for the heat stability, RCT and casein content traits under the L-BPLS model for one test fold of the MIR milk spectral data; 40 random test-set milk samples are given in increasing order of respective trait quantities. True trait values and predictions from standard PLS are also provided.}
	\label{fig:MIRmilk_PredIntervals_Units}
\end{figure}

The technological traits of both heat stability and RCT, as well as a protein trait related to the casein content of the milk samples were jointly predicted using the proposed BPLS methods from the milk MIR spectral data (see Section \ref{sec:mir_milk_trait}). As shown in Table \ref{tab:mir_milkHSRC31}, the traits are predicted with similar accuracy under the proposed BPLS models and existing methods, with  L-BPLS offering the most accurate point predictions for all three traits; further prediction accuracy results comparing BPLS regression to Bayesian factor analysis approaches appear in Appendix F.1 of the Supplementary Material.  While accurate milk trait prediction is important, BPLS models offer additional utility in the form of prediction intervals as seen in Figure \ref{fig:MIRmilk_PredIntervals_Units}. The benefit of using the proposed BPLS methodology is exemplified in the heat stability predictions (Figure \ref{fig:MIRmilk_PredIntervals_Units}, left): where PLS produces very inaccurate point predictions for milk samples $26$ and $40$, the predictions from our model are substantially closer to the true trait value. Further, when the BPLS predictions are farther from the true trait values, the uncertainty is conveyed through the provision of coherent prediction intervals (e.g.~milk sample $30$).  

The quantification of the uncertainty associated with the predictions can be particularly useful for milk processors segregating incoming milk streams for different value markets; e.g.\ cheese, butter or infant formula. Such segregation may prevent the undesirable introduction of inferior milk streams into a pool intended for a specific value product. The provision of reliable prediction intervals can give processors confidence that a particular standard has been met: if a given point prediction is above an acceptable threshold but there is substantial posterior predictive mass below it, the product is likely to fail to meet the standard. In such cases, the processor may commission secondary analyses using a gold standard method on a sample of the milk pool e.g.\ measuring heat stability using an oil bath \cite[][]{DuHu2020}.  Additionally, stringent food labelling  policies require that the properties (e.g.\ nutritional values) of a given product reside within some acceptable margin of error. The provision of prediction intervals here would give processors confidence that a given product's stated properties are sufficiently likely to correspond to what is being delivered. The BPLS approach is therefore beneficial to stakeholders in dairy production, aiding more informed management decisions by taking into account the uncertainty in the predictions, particularly when incorrect decisions may have health and/or financial repercussions.

\begin{table}[tb]
	\caption{Accuracy of predictions of the standardised heat stability, casein content and RCT traits from the milk MIR spectral data using each of the proposed BPLS and existing methods. Estimates are given as ``mean (standard deviation)'' of the RMSEP over the four test folds, with bold font indicating optimal performance. $^*$Due to numerical instability, $\widehat{\ttQ}_y$ was picked manually.}
	\label{tab:mir_milkHSRC31}
	\centering

	\begin{tabular}{l|cc|cc|cc}

	\hline
	& \multicolumn{2}{c}{\textbf{Heat stability}} &  \multicolumn{2}{|c}{\textbf{Casein content}} &\multicolumn{2}{|c}{\textbf{RCT}}  \\
	& RMSEP&$\widehat{\ttQ}_y$ &RMSEP&$\widehat{\ttQ}_y$& RMSEP&$\widehat{\ttQ}_y$ \\
	\hline
BBPLS & 0.875 (0.172) & 6.5 (2.5) & 0.376 (0.027) & 6.8 (1.5) & 0.815 (0.139) & 5.5 (1.9) \\ 
  ss-BPLS & 0.874 (0.164) & 4.3 (0.5) & 0.376 (0.031) & 6.3 (1.3) & 0.813 (0.141) & 5.3 (1.0) \\ 
  L-BPLS & \textbf{0.823} (0.156) & 15.8 (1.0) & \textbf{0.360} (0.041) & 15.8 (2.1) & \textbf{0.782} (0.126) & 18.3 (3.0) \\ \hline
  PLS & 0.896 (0.117) & 5.5 (0.6) & 0.395 (0.028) & 4.5 (0.6) & 0.863 (0.141) & 5.5 (1.7) \\ 
  PLS-1s & 0.849 (0.125) & 7.8 (1.5) & 0.397 (0.022) & 4.5 (0.6) & 0.810 (0.105) & 7.3 (1.3) \\ 
  sPLS & 0.879 (0.173) & 14.5 (4.0) & 0.376 (0.044) & 8.5 (4.4) & 0.801 (0.128) & 14.3 (3.5) \\ 
  PCR & 0.877 (0.112) & 11.3 (1.9) & 0.402 (0.024) & 4.8 (1.5) & 0.824 (0.089) & 14.3 (7.2) \\ 
  LASSO & 0.846 (0.121) & -- & 0.406 (0.024) & -- & 0.845 (0.125) & -- \\ 
  RIDGE & 0.992 (0.145) & -- & 0.446 (0.027) & -- & 0.925 (0.108) & -- \\ 
  PPLS &0.987 (0.127) & $^*$2.0 (0.0) & 0.682 (0.082) & $^*$2.0 (0.0) & 0.970 (0.095) & $^*$2.0 (0.0) \\
   \hline
	\end{tabular}
\end{table}




As is good statistical practice, to validate the reported prediction intervals the prediction interval coverage is reported in Table \ref{tab:mir_coverage} for each of the three BPLS methods. Here, L-BPLS interval widths are narrower for the heat stability trait compared to the other two methods---this could be due to the additional bias introduced by the LASSO-type penalty. All three methods have slightly overconfident intervals for the RCT trait, with BPLS exhibiting the strongest bias, suggesting potentially inadequate model fit; the coverage could be further improved by adjusting the variable transformation.
It is noteworthy that the estimated prediction interval coverage for casein was very close to the target $95\%$. Casein constitutes approximately 80\% of total milk protein \citep{FaJi2004} and the total protein content of milk can be predicted very accurately with standard methods \citep{BoDi2011}.
For all three traits, the Gaussianity assumption appears appropriate, particularly for casein. The slightly too narrow prediction intervals for heat stability and RCT suggest a possibly more nuanced relationship between these technological traits of milk and the MIR spectra. Furthermore, there can ambiguity about the ground-truth measurements of the traits themselves (e.g.\ measuring the heat stability of milk using an oil bath is arguably subjective), naturally limiting the accuracy of any statistical model.

Further analyses of the MIR data where two quarters of data are used for training and the remaining two are used for testing can be found in Appendix F.2 of the Supplementary Material.

\begin{table}[tb]
\centering
\caption{Medians, across four data folds, of the $95\%$ prediction interval coverage estimates from the milk MIR spectral data. Estimates are based on the proportion of true test sample traits contained within the predicted intervals for the three BPLS models.}\label{tab:mir_coverage}
\begin{tabular}{l|c|c|c}
        & \textbf{Heat stability} & \textbf{Casein content} & \textbf{RCT} \\ \hline
BPLS    &  0.940  & 0.951       &  0.924   \\
ss-BPLS &  0.946  & 0.957       &  0.940   \\
L-BPLS  &  0.912  & 0.951       &  0.940  
\end{tabular}
\end{table}

\subsection{Predicting pH from SERS milk spectral data} 
\label{sec:data_examples_SERS}

As described in Section \ref{sec:sers_milk}, interest lies in predicting the pH of milk samples from their SERS intensity data. The suite of BPLS models and existing methods were used to predict the pH, with the BPLS models offering a clear improvement over the competitors (see Table \ref{tab:sers_milk}); additional performance comparisons between BPLS models and Bayesian factor analysis approaches are given in Appendix \ref{app:bfa} of the Supplementary Material. The best-performing method in terms of RMSEP is again L-BPLS, which results in a 34.7\% decrease in the error compared to the popular sPLS method. This is considered a particularly challenging data setting as $N = 11 \ll \mathtt{P} = 1733$; the BPLS methods demonstrated particularly strong performance in such settings in Section \ref{sec:numerical_experiments}.

\begin{figure}[tb]
	\centering
 \includegraphics[width = 0.32\textwidth]{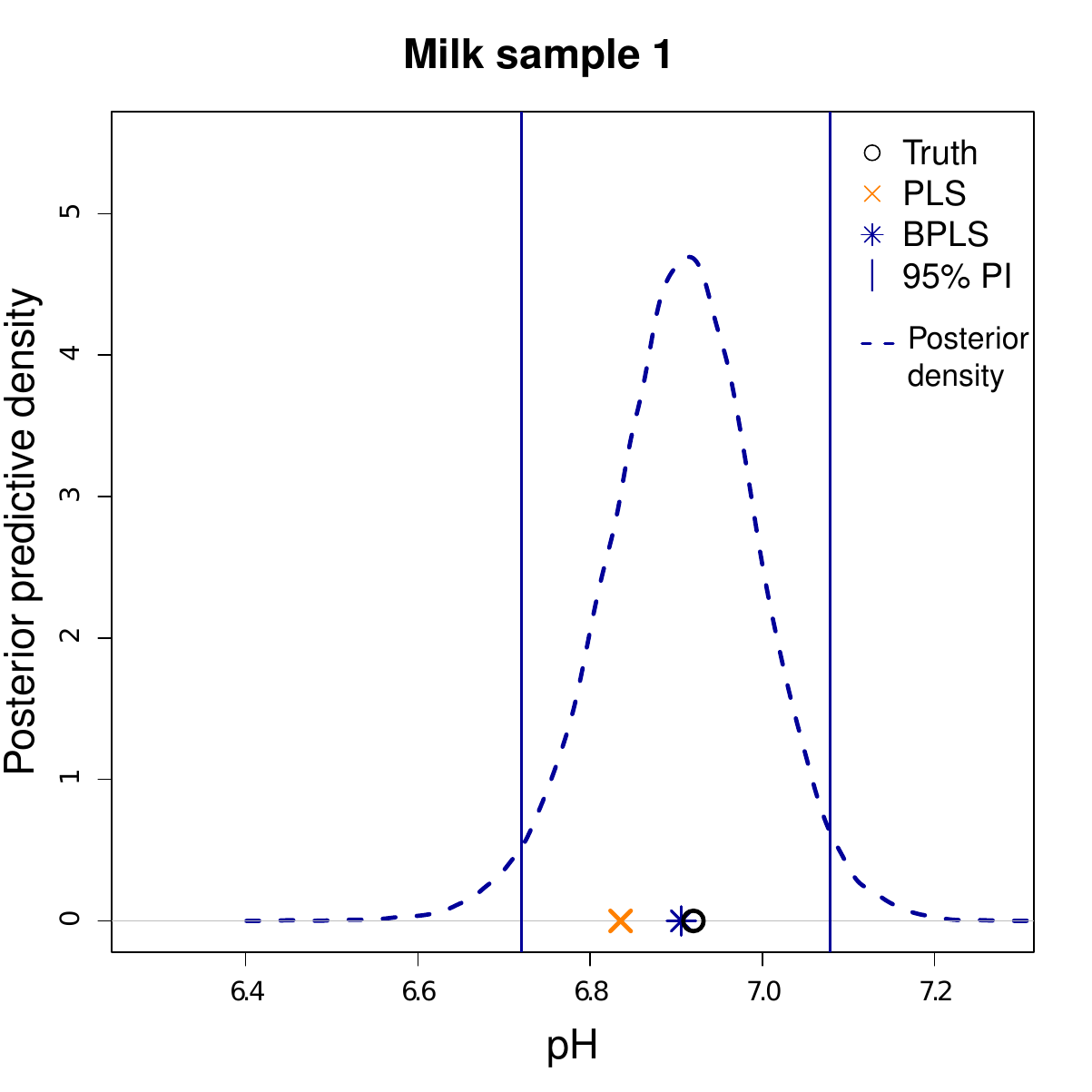}
 \includegraphics[width = 0.32\textwidth]{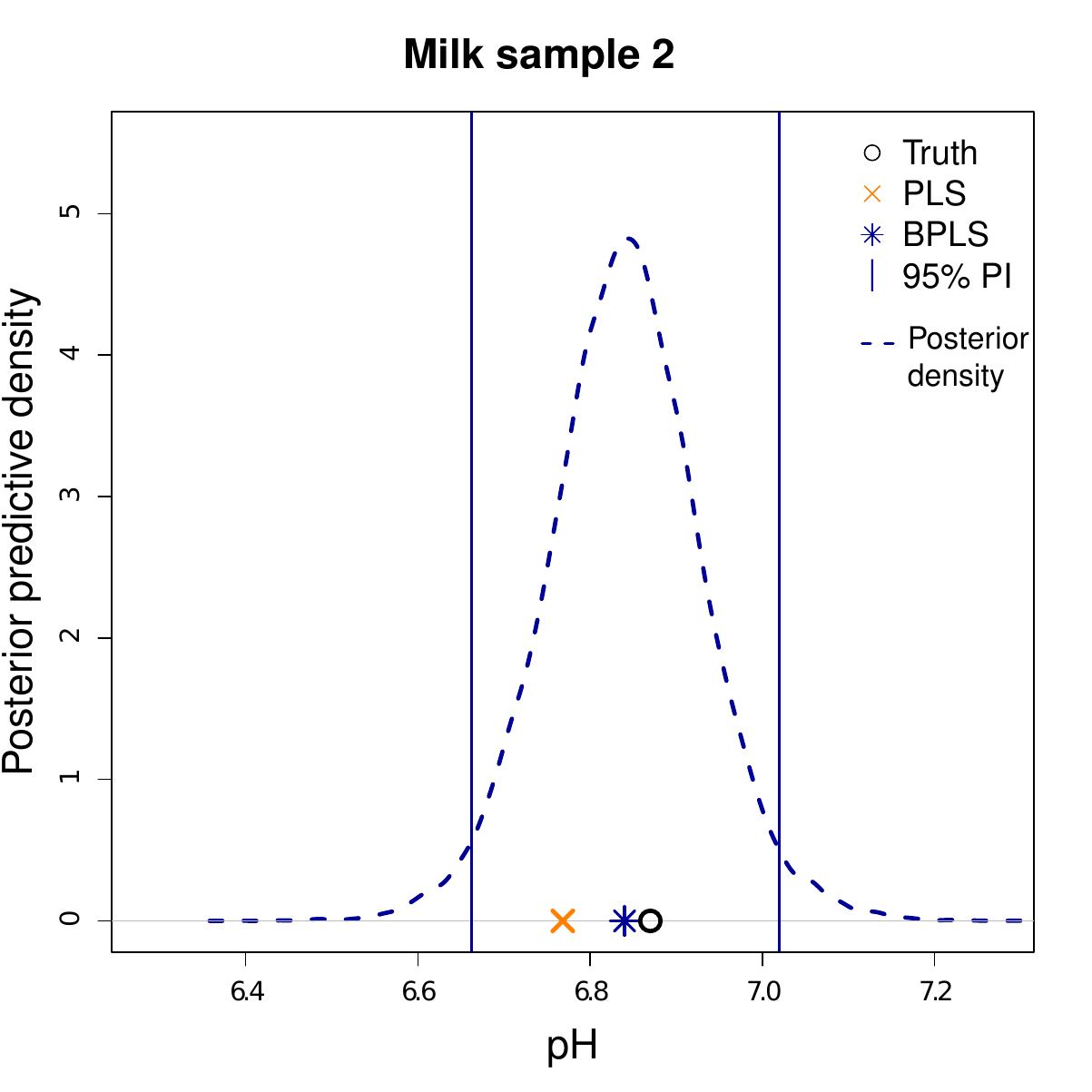}
 \includegraphics[width = 0.32\textwidth]{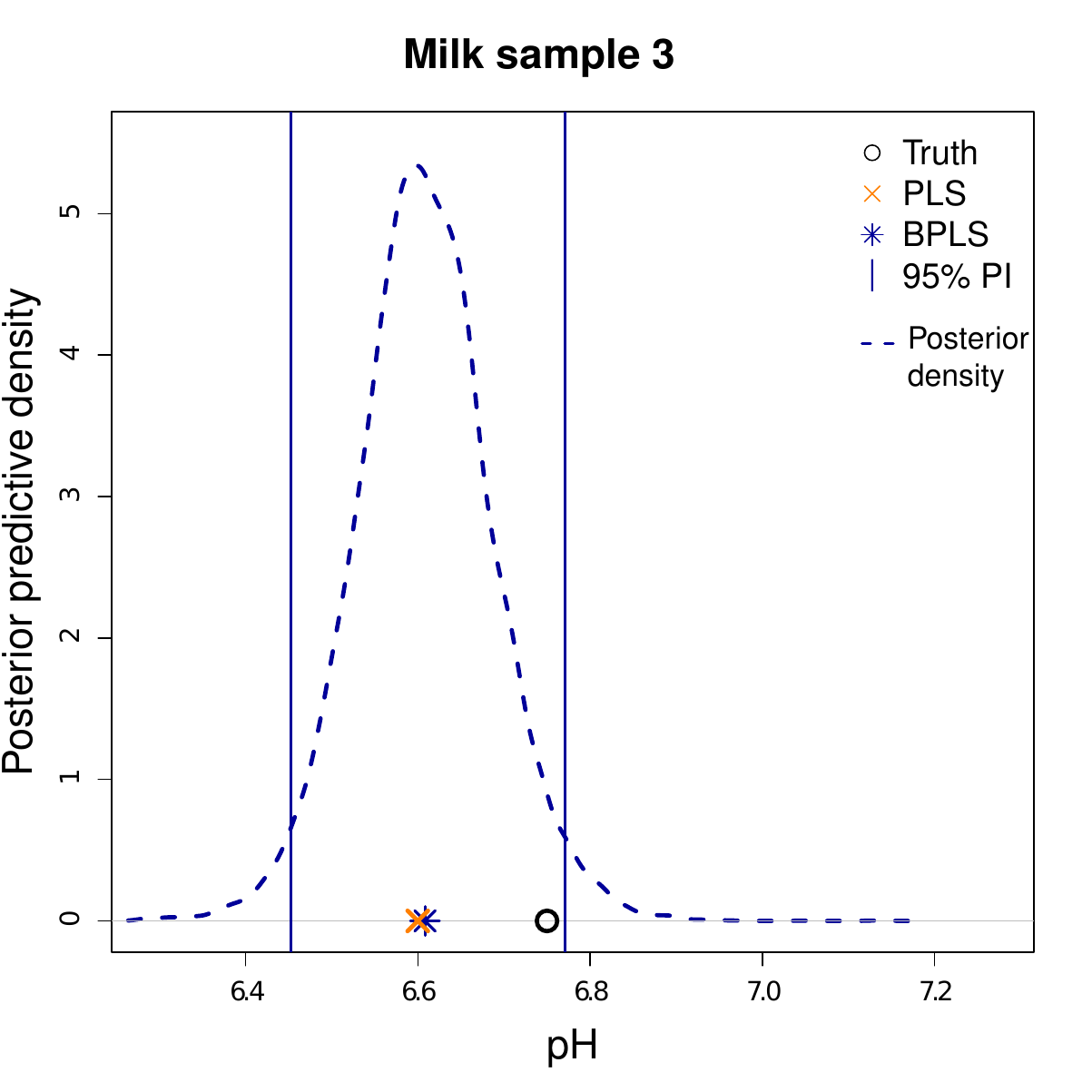}
	\caption{Posterior predictive densities (blue, dashed) of pH for three test milk samples (with posterior predictive means and $95\%$ prediction intervals marked) obtained from the L-BPLS model from one test fold of the milk pH SERS dataset, along with standard PLS predictions.}
	\label{fig:SERSmilk_PredIntervals}
\end{figure}

Figure \ref{fig:SERSmilk_PredIntervals} illustrates the posterior predictive densities of pH, which are naturally available under the proposed L-BPLS model, for three milk samples from one test fold of the SERS data; posterior predictive means and 95\% prediction intervals are highlighted, as is the prediction from PLS regression. The L-BPLS predictions were more accurate than the PLS regression approach and, even when the predictions from L-BPLS were not close to the gold standard value, the latter resided within the 95\% prediction interval (e.g.~milk sample 3, Figure \ref{fig:SERSmilk_PredIntervals} (right)). 

Access to posterior predictive distributions for milk sample traits enables the end-users (here milk producers or processors) to make probabilistic statements about the milk pH, especially in relation to milk spoilage and animal health. A point prediction indicating a low pH could be an indication of mastitis \citep{KaMe2019}. However, the presence of high uncertainty in the form of a wide prediction interval would dissuade immediate remedial action, minimising unnecessary medical interventions, through the use of antibiotics, with downstream repercussions on anti-microbial resistance. Instead, the end user may pay more attention to the animal or may perform additional tests \citep[e.g.\ California mastitis test;][]{ScNo1957}.

\begin{table}[tb]
	\caption{Accuracy of pH predictions from the milk SERS data under the BPLS and existing methods. Estimates are given as `mean (standard deviation)' over the four test folds, with bold font indicating optimal performance. $^*$Due to numerical instability, $\widehat{\ttQ}_y$ was picked manually.}
	\label{tab:sers_milk}
	\centering

	\begin{tabular}{l|cc}
	\hline

	\hline
	&\multicolumn{2}{c}{\textbf{pH}}  \\
	&RMSEP&$\widehat{\ttQ}_y$  \\
	\hline
	BPLS & 0.627 (0.248) & 6.0 (1.4) \\ 
	ss-BPLS & 0.644 (0.272) & 4.3 (1.0) \\ 
	L-BPLS & \textbf{0.614} (0.232) & 5.5 (0.6) \\ \hline
	PLS & 0.959 (0.330) & 1.3 (0.5) \\ 
	PLS-1s & 0.959 (0.330) & $^*$0.3 (1.5) \\ 
	sPLS & 0.941 (0.293) & 4.0 (1.6) \\ 
	PCR & 1.024 (0.374) & 0.3 (0.5) \\ 
	LASSO & 0.985 (0.471) & -- \\ 
	RIDGE & 0.946 (0.453) & -- \\ 
	PPLS & 0.783 (0.573)& $^*$1.0 (0.0)\\
	\hline

	\hline
	\end{tabular}
\end{table}


\section{Discussion} 
\label{sec:discussion}
We address the problem statement in the global agri-food industry in relation to the simultaneous prediction of multivariate phenotype vectors  using high-dimensional spectral data, specifically in the context of dairy production. There is a need for accurate and informative milk trait prediction methodologies which can utilise the spectral data that are routinely generated, as in the Irish agri-food sector. To achieve this, we introduce a probabilistic Gaussian latent-variable model, BPLS, which emulates the popular partial least squares regression algorithm, along with two variants of the BPLS model inspired by sparse regression methods. Posterior inference for the suite of BPLS methods was carried out through a MCMC  algorithm. While numerical experiments explored the performance of the BPLS models, improved prediction accuracy of milk traits from the motivating MIR and SERS spectral data sets was achieved under the proposed BPLS models compared to the industry-standard PLS approaches. Importantly, the BPLS methods bestow additional benefits such as the coherent provision of prediction intervals for the traits which can assist in informed decision-making by milk producers and processors. They obviate the need for the user to select a number of latent components, and facilitate the prediction of traits in a multivariate fashion. In preliminary analyses, when separate BPLS models were fitted to individual milk traits, the overall predictions worsened indicating a clear benefit from the multivariate modelling approach which is facilitated by the BPLS's basis in a statistical model. This is in contrast to multivariate PLS-type models where picking $\ttQ$ can be particularly challenging---one may be forced to overfit for one trait to compensate for another; this is illustrated in Appendix \ref{app:multivariate} of the Supplementary Material. The BPLS methods also performed well in the SERS data setting which is a ``small $N$, large $\ttP$'' scenario; the regularisation provided through the use of Bayesian priors allowed inference even in extreme settings where the number of parameters was over a hundred times larger than the number of milk samples. In summary, treating PLS regression from a probabilistic viewpoint, drawing on Bayesian nonparametric priors to obviate model selection challenges and incorporating sparse regression strategies results in a single, coherent BPLS regression modelling framework for prediction of dairy traits from spectral data, with uncertainty inherently provided.

Based on the thorough empirical comparisons, we found that the LASSO variant of BPLS had the most reliable prediction accuracy for the datasets considered here. The sparsity structure arising from the combination of Bayesian LASSO and shrinkage priors resulted in a model with very accurate posterior predictive means. Whilst L-BPLS was also robust in estimating the predictive variance in the synthetic datasets, as indicated by the coverage probability results, it underperformed in the MIR spectral data setting where ss-BPLS achieved more desirable coverage. As is often the case, no single model works best in all situations. When working with spectral data, L-BPLS may be more favourable if point prediction accuracy is more important than robust uncertainty quantification, but if the opposite is the case, ss-BPLS may be the more robust choice. Based on these results, the recommendation would be to fit the two models to the given spectral data problem and validate for a given objective (e.g.~point prediction or coverage probability) prior to deployment.

While technological and protein milk traits were considered here, the BPLS framework does not rely on a particular type of milk trait. Indeed, BPLS methods could be used for predictions of quantities not directly related to milk such as cattle methane output \citep{CoVa2022}, body energy status of cattle \citep{FrGo2023}, cattle reproductive performance \citep{ToPe2021} or presence of disease \citep{DeBr2020}. Equally, the BPLS models are not confined to mid-infrared or surface-enhanced Raman spectra from milk samples and could be applied to problems involving ultraviolet, visible or near-infrared spectral data \citep[e.g.][]{AePo2011,BeZa2019}. The underpinning of BPLS by a coherent probabilistic model would aid researchers in predicting a wide range of traits in the agri-food domain. 

The BPLS models as introduced, and many of their existing competitors, assume independent and identically distributed observations, conditional on the model parameters. This can be an incorrect modelling assumption, particularly for problems encountered in the agri-food context. It is not uncommon for agri-food datasets to contain multiple samples relating to a single source, e.g.~in milk MIR spectral datasets, several milk samples typically come from one cow. The BPLS's probabilistic formulation elegantly lends itself to facilitate modifications to appropriately model such dependency structures, e.g.~it could allow for a hierarchical structure with random effects, appropriately accounting for first-order interactions. Such extensions are less feasible in existing, non-probabilistic methods such as PLS. 

Given the motivating application, in this work, we focused on predicting real-valued traits, however, a great deal of interest lies in problems involving the prediction of binary responses from spectral data, e.g.~predicting a cow's pregnancy status \citep{BrWe2021} or diet type \citep{CoMa2021}. Generalised linear model-oriented extensions to the BPLS methods would naturally and appropriately allow for such categorical responses, with random effects again permitting faithful modelling of repeated measures data. 
Indeed in \cite{FrGo2021}, separate PLS methods were used to predict the RCT and heat stability traits from milk samples; samples were subsequently classified into groups based on high and low levels of the traits with PLS discriminant analysis \citep{BaRa2003} relied on for the latter task. Both tasks could be simultaneously achieved using the same posterior predictive distributions from a single BPLS model fit.

Given the practical utility of the developed methods in dairy production, one potential concern is the scalability of the BPLS framework as the number of variables, $\ttP$, grows. A naive $\textsc{R}$ implementation of the Gibbs algorithm was sufficiently computationally efficient for the MIR and SERS data, with run times of 18 and 30 minutes respectively on a Dell Latitude 5430 laptop, equipped with a 4.40 GHz Intel core i5-1235U processor and 32 GB of RAM. In practice, regression models would not require frequent refitting as the collection of traits can be costly and takes time. More sophisticated future models with complex hierarchical structures would likely require approximate inference schemes  to alleviate the computational burden that comes with the growing parameter space \citep[e.g.][]{HaAv2023}; however, caution should be taken to ensure a controlled approximation error.
Relatedly, the BPLS framework does not strictly depend on the infinitely-dimensional nonparametric shrinkage prior. A finite-dimensional variant could be employed by performing model selection based on information criteria such as BIC-MCMC \citep{Fruh2011}; albeit, fitting multiple models would come at a steep computational cost.

\section{Acknowledgements}
The authors would like to thank the members of the VistaMilk Science Foundation Ireland (SFI) Research Center, and in particular the members of the VistaMilk Spectroscopy Working Group, for discussions that contributed to this work. The authors are grateful to the editors and the two anonymous reviewers for comments and suggestions that have materially improved the article.\\
This publication has emanated from research conducted with the financial support of Science Foundation Ireland (SFI) and the Department of Agriculture, Food and Marine on behalf of the Government of Ireland under grant number (16/RC/3835) and the SFI Insight Research Centre under grant number (SFI/12/RC/2289\_P2).
\bibliography{ppca_refs}
\bibliographystyle{apalike}

\newpage
\appendix
\setcounter{page}{1}
\setcounter{figure}{0}    
\setcounter{table}{0} 
\section*{Supplementary material}
This file contains the technical appendix for ``Predicting milk traits from spectral data using Bayesian probabilistic partial least squares regression''.

\begin{itemize}
    \item Appendix \ref{app:resp_transform} outlines the transformation used on positive-valued response to map them to real numbers.
    \item Appendix \ref{app:mcmc_details} gives the details of the Markov chain Monte Carlo algorithms used for inference in the Bayesian partial least squares regression models.
    \item Appendix \ref{app:posterior_shrinkage} illustrates how the fitted Bayesian partial least squares regression model achieves desired shrinkage.
    \item Appendix \ref{app:posterior_predictive} outlines a computationally efficient method for approximating the posterior predictive distribution of a response from a set parameter posterior samples, as well as a comparison using predictions conditional on the posterior parameter modes.
    \item Appendix \ref{app:sim_results} provides additional numerical comparison results on synthetic data
    \item Appendix \ref{app:additional_data} provides additional results from analyses of the motivating mid-infrared milk spectral data and the benchmark near-infrared grain mash spectral.  
\end{itemize}

\section{Response transformation}
\label{app:resp_transform}
\begin{figure}[h]
    \centering
    \includegraphics[width = 0.7\textwidth]{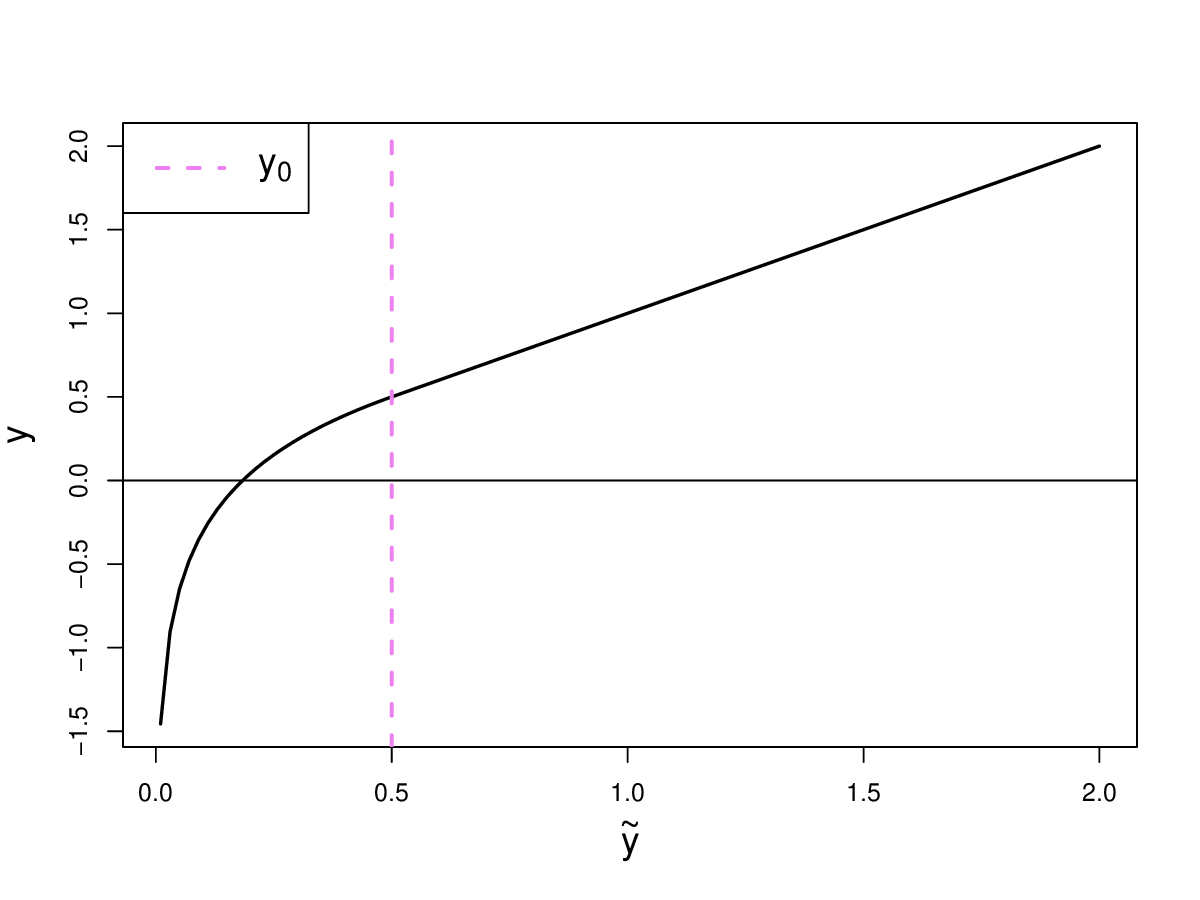}
    \caption{Non-linear transformation $f:[0,\infty)\rightarrow\Real$ of the response, given by \eqref{eqn:resp_transform}. The original response variable $\tilde{y}$ is mapped to a real-valued $y$. Here, $y_0$ is arbitrarily set to $0.5$.}
    \label{fig:resp_transform}
\end{figure}

Interest lies in predicting traits that in reality are strictly positive; that is, the response variable on the original scale $\tilde{y}$ is such that $\tilde{y}>0$. Early empirical results showed that the typically standard logarithm transform of the response ($y=\log\tilde{y}$) can negatively impact the accuracy of the predictions in the proposed methods as well as competitor ones.On the original scale of the response, this corresponds to the presence of multiplicative noise. To mitigate this issue we use the following transform which ensures the correct support of the response variable:
\begin{align}
    f(\tilde{y})=\begin{cases}
    y_0\left(\log\left(\tilde{y}/y_0\right)+1\right)&\tilde{y}<y_0\\
    \tilde{y}&\tilde{y}\geq y_0\end{cases},\label{eqn:resp_transform}
\end{align}
where $y_0$ is a user-specified changepoint; we found that setting $y_0$ to either the minimum observed value or the sample mean minus two sample standard deviations gave valid predictions at a negligible change to the prediction accuracy; the original data remain largely unaffected as the transformation is the identity for the bulk of the observations. It can be verified that $f$ is differentiable on $[0,\infty)$. Figure \ref{fig:resp_transform} illustrates the proposed transformation where $y_0=0.5$. After applying the $f$ transformation the resulting variables are standardised to have zero mean and unit variance prior to model fitting to comply with the BPLS model formulation; this standardisation is then reversed after prediction from the model to ensure practically valid predictions.





\section{Posterior sampling for BPLS} 
\label{app:mcmc_details}

Here, we list all the conditional distributions required for the Gibbs algorithm for the posterior sampling BPLS model parameters. Recall that the data are made up of $N$ samples, each with $\ttP$ predictors and $\ttR$ responses. We let $\bmx^{(p)}$ be the $p\Th$ column of $\bfX$ ($p=1,\ldots,\ttP)$, and $\bmy^{(r)}$ be the $r\Th$ column of $\bfY$ ($r=1,\ldots,\ttR$). With a slight abuse of notation we use ``$\ldots$'' to denote all other variables involved in the inference scheme. For efficient sampling, we sample the loading matrix entries by whole rows. The below assumes a particular truncation $\ttQ^*$ of the infinite latent dimension $\ttQ$, according to the advice given in Section 3.2 of the main paper.
\begin{align*}
	\bfz_n|\ldots&\sim\normal(S\left(W^\top\Sigma\inv\bfx_n +  C^\top \Psi\inv\bfy_n\right),S),~~n=1,\ldots,N;\\
	\bfw_p|\ldots&\sim\normal\left(\psi_p^{-2}\widetilde{\Delta}_p\bfZ^\top \bmx^{(p)},\widetilde{\Delta}_p\right),~~p=1,\ldots,\ttP;\\
	\bfc_r|\ldots&\sim\normal\left(\sigma_r^{-2}\widetilde{\Omega}_r \bfZ^\top \bmy^{(r)},\widetilde{\Omega}_j\right),~~r=1,\ldots,\ttR;\\
	\sigma_p^{-2}|...&\sim\gamdist\left(A_\sigma+\frac{N}{2},B_\sigma + \frac{1}{2}\left\|\bmx^{(p)}-\bfw_p^{\top} \bfZ\right\|^2\right),~~p=1,\ldots,\ttP;\\
	\psi_r^{-2}|...&\sim\gamdist\left(A_\psi+\frac{N}{2},B_\psi + \frac{1}{2}\left\|\bmy^{(r)}-\bfc_r^{\top}\bfZ\right\|^2\right)~~r=1,\ldots,\ttR;\\
	\phi_{pq}|...&\sim\gamdist\left(\nu^w_1+\frac{1}{2},\nu^w_2+\frac{w_{pq}^2\tau_q}{2}\right),~~ p=1,\ldots,\ttP,~q=1,\ldots,\ttQ^*;\\
	\dot\phi_{rq}|...&\sim\gamdist\left(\nu^c_1+\frac{1}{2},\nu^c_2+\frac{c_{rq}^2\tau_q}{2}\right)~~r=1,\ldots,\ttR,~~q=1,\ldots,\ttQ^*;\\
	\delta_k|...&\sim\gamdist^*\left(\alpha+\frac{\ttP(\ttQ^*-k+1)}{2},\beta + \frac{1}{2}\sum_{q=k}^\ttQ\tau^{(k)}_q\left(\sum_{p=1}^\ttP\phi_{pq}w_{pq}^2+\sum_{r=1}^\ttP\dot\phi_{rq}c_{rq}^2\right)\right)\\
	&~~~~~~k=1,\ldots,\ttQ^*;
\end{align*}
where
\begin{align*}
	S &= \left(\Id{\ttQ}+W^\top\Sigma\inv W +  C^\top \Psi\inv C \right)\inv,\\
	\widetilde{\Delta}_p &= \left(\Delta_p\inv+\psi_p^{-2}\bfZ^\top\bfZ\right)\inv,\\
	\Delta_p &= \diag(\tau_1\phi_{p1},\ldots,\tau_q\phi_{pq},\ldots),\\
	\widetilde{\Omega}_r&= \left(\Omega_r\inv + \sigma_r^{-2}\bfZ^{\top}\bfZ \right)\inv,\\
	\Omega_p &= \diag(\tau_1\dot\phi_{r1},\ldots,\tau_q\dot\phi_{rq},\ldots),\\
	\tau^{(k)}_q &= \prod_{l=1}^{\ttQ^*}\delta_l^{1-\Ind{l=k}}, \delta_1 = 1.
\end{align*}

If the noise is assumed to be isotropic, that is, $\sigma^2_p\equiv\sigma^2~\forall p$, then
\begin{equation*}
	\sigma^{-2}\sim\gamdist(A_\sigma,B_\sigma)\implies\sigma^{-2}|...\sim\gamdist\left(A_\sigma+\frac{N\times\ttP}{2},B_\sigma + \frac{1}{2}\sum_{p=1}^\ttP\left\|\bmx^{(p)}-\bfw_p^{\top} \bfZ\right\|^2\right)
\end{equation*}
and similarly,  for $\psi^2_r\equiv\psi^2~\forall r$,
\begin{equation*}
	\psi^{-2}\sim\gamdist(A_\psi,B_\psi)\implies\psi^{-2}|...\sim\gamdist\left(A_\psi+\frac{N\times\ttR}{2},B_\psi + \frac{1}{2}\sum_{r=1}^\ttR\left\|\bmy^{(r)}-\bfc_r^{\top}\bfZ\right\|^2\right).
\end{equation*}

\subsection{Additional posterior sampling details for ss-BPLS} 
\label{app:gibbs_ss_bpls}
Recall that in the spike-and-slab variant of the BPLS model the generative model for the response is $\bfy = CB\bfz + \bmeta$, where $B$ is a binary diagonal matrix. \emph{A priori}, we let the diagonal entries be Bernoulli random variables, each with success probability $p_0$. We outline a Metropolis-Hastings (MH) update for this random matrix; it requires the calculation of the posterior odds of the current $B$ and the proposal $B'$. We only update $B$ one element at a time, using notation $D = B'-B$, where $D$ is a zero everywhere apart from the $i,i$ entry, which itself is equal to $\gamma\in\{-1,1\}$. For each component $q=1,...,\ttQ^*$, consider the deterministic proposal $f_q(B'|B) = \Ind{B' = B+D}$, where $[D]_{i,i} = \gamma = 1-2[B]_{i,i}$ for $i=q$, and $0$ otherwise; clearly, the proposal is symmetric. The combination of $\ttQ^*$ such updates (one for each variable) results in a kernel which preserves $\pi$. The MH acceptance probability is then 
\begin{align*}
	\alpha(B,B') &= 1\wedge\frac{\pi(B'|...)}{\pi(B|...)}\\
						&= 1\wedge\frac{\p(\bfY|B',...)\pi_0(B')}{\p(\bfY|B,...)\pi_0(B)}.
\end{align*}
To calculate the ratio of the likelihood terms, we first note that for a single observation
\begin{align*}
	\p(\bfy_n|B',...)
	&=(2\pi)^{-\frac{r}{2}}|\Psi|^{-\frac{1}{2}} \exp\left(-\frac{1}{2}\left(\bfy_n - C\widetilde{B}\bfz_n\right)^\top\Psi\inv\left(\bfy_n - C\widetilde{B}\bfz_n\right)\right)\\
	&=(2\pi)^{-\frac{r}{2}}|\Psi|^{-\frac{1}{2}}  \exp\left(-\frac{1}{2}\left(\bfy_n - C(B+D)\bfz_n\right)^\top\Psi\inv\left(\bfy_n - C(B+D)\bfz_n\right)\right)\\
	&=(2\pi)^{-\frac{r}{2}}|\Psi|^{-\frac{1}{2}}  \exp\left(-\frac{1}{2}\left(\bfy_n - CB\bfz_n\right)^\top\Psi\inv\left(\bfy_n - CB\bfz_n\right) + \bfz_n^\top D^\top C^\top\Psi\inv C D\bfz_n\right.\\
	&~~~~~~~~~~~~~~~~~~~~~~~~~~~~~~~~~~~~~~~~~~\left. - 2\bfz^\top D^\top C^\top\Psi\inv\left(\bfy_n - CB\bfz_n\right) \right)\\
	&= \p(\bfy|B,...)\times\e^{-\frac{1}{2}\upsilon_n}.
\end{align*}
Thus the product over the whole dataset follows $\p(\bfY|B',...)=\p(\bfY|B,...)\times\e^{-\frac{1}{2}\sum_{n=1}^N\upsilon_n}$. Since $CD\bfz = \gamma z_q\bmc^{(q)}$, where $\bmc^{(q)}$ is the $q\Th$ column of $C$, the likelihood contribution for a single observation $n$ ($=1,...,N$) involves the quantity
\begin{align*}
	\upsilon_n = z_{nq}^2\bmc^{(q)\top}\Psi\inv\bmc^{(q)} - 2\gamma z_{nq}\bmc^{(q)\top}\Psi\inv\left(\bfy_n - CB\bfz_n\right).
\end{align*}
Taking the sum over the observation indices
\begin{align*}
	\sum_{n=1}^N\upsilon_i 
	&= \sum_{n=1}^n z_{nq}^2\bmc^{(q)\top}\Psi\inv\bmc^{(q)} - 2\gamma z_{nq}\bmc^{(q)\top}\Psi\inv\left(\bfy_n - CB\bfz_n\right)\\
	&=\bmc^{(q)\top}\Psi\inv\bmc^{(q)}\left(\sum_{n=1}^N z_{iq}^2\right)-2\gamma \bmc^{(q)\top}\Psi\inv\left(\sum_{n=1}^N z_{nq}\left(\bfy_n - CB\bfz_n\right)\right)\\
	&=\bmc^{(q)\top}\Psi\inv\bmc^{(q)} \bmz^{(q)\top}\bmz^{(q)}-2\gamma \bmc^{(q)\top}\Psi\inv\left(\bfY-\bfZ B C^\top\right)^\top\bmz^{(q)}\\
	&=\bmc^{(q)\top}\Psi\inv\left[\bmc^{(q)} \bmz^{(q)\top}-2\gamma \left(\bfY-\bfZ B C^\top\right)^\top\right]\bmz^{(q)}.
\end{align*}

Thus, with prior probability of slab $p_0$, the acceptance probability for the move becomes
\[
	\alpha(B,B') = 1\wedge \e^{-\frac{1}{2}\bmc^{(q)\top}\Psi\inv\left[\bmc^{(q)} \bmz^{(q)\top}-2\gamma \left(\bfY-\bfZ B C^\top\right)^\top\right]\bmz^{(q)}}\left(\frac{p_0}{1-p_0}\right)^\gamma.
\]
The probability $p_0$ is assigned a conjugate $\betadist(\alpha_S,\beta_S)$ prior. Thus, with a finite dimension truncation $\ttQ^*$, the conditional posterior is $p_0|\ldots\sim\betadist\left(\alpha_S+\|\bfb\|_1 ,\beta_S+\ttQ^*-\|\bfb\|_1\right)$, where $\bfb$ is a vector of the diagonal entries of $B$.


\subsection{Additional posterior sampling details for L-BPLS} 
\label{app:gibbs_l_bpls}
In the Bayesian LASSO variant of BPLS, we now use the prior structure
\begin{align*}
 	c_{rq}|\dot{\phi}_{rq},\tau_q\sim\normal(0,\dot{\phi}_{rq}\inv\tau_q\inv),~~\dot{\phi}_{rq}\inv\sim\Exp\left(\lambda^2/2\right),~r=1,\ldots,\ttR,~q=1,2,\ldots\, , 
 \end{align*}
 where $\bmtau = (\tau_1,\tau_2,\ldots)$ is the same shrinkage prior as given in Section 3.2 of the main paper. Exact Gibbs sampling can be carried out in this scenario as the conditional of $\dot{\phi}_{rq}$ has an inverse-Gaussian distribution,
\begin{equation}
	\pi(\dot{\phi}_{rq}|...) = \frac{\lambda}{\sqrt{2\pi}}\dot{\phi}^{3/2}_{rq}\exp\left(-\frac{\lambda^2(\dot{\phi}_{rq}-\mu_{rq})^2}{2\mu_{rq}^2\dot{\phi}_{rq}}\right),~~\mbox{where}~~\mu_{rq} = \sqrt{\frac{\lambda^2}{\tau_q c_{rq}^2}}.
\end{equation}

Tractable inference on $\lambda$ can be carried by assigning $\lambda^2$ a conjugate gamma prior; specifically,
\begin{equation}
	\lambda^2\sim\gamdist\left(A_\lambda+\ttR\times\ttQ^*,B_\lambda+\frac{1}{2}\sum_{r,q}\dot{\phi}_{rq}\inv\right).
\end{equation}


\subsection{MCMC initialisation} 
\label{app:mcmc_initialisation}
To avoid a completely ``cold start'' in the Gibbs sampling algorithm, the chain is initialised according to the following set of guidelines:
\begin{enumerate}
	\item Perform PCA on $\bfX$ and identify $\ttQ^*$ as the number of principal components which explains at least $0.99$ of the observed variation.
	\item Set the columns of $W$ to be the first $\ttQ^*$ loading vectors.
	\item Set all $\sigma_p^2$, $p=1,\ldots,\ttP$, to be equal to average of the eigenvalues corresponding to the remaining loading vectors.
	\item Set $\bfZ = \bfX\Sigma\inv W S_z$ (conditional mean) where $S_z = \left(\Id{\ttQ}+ W^\top\Sigma\inv W\right)\inv$; and set $C^\top = \left(\bfZ^\top\bfZ\right)\inv \bfZ^\top\bfY$ (least squares estimate)
	\item Initialise all remaining parameters from respective prior distributions.
	\item Adjust $\ttQ^*$ based on a warm-up run.
\end{enumerate}

\section{Posterior distribution of shrinkage variables}\label{app:posterior_shrinkage}
\begin{figure}[tb]
	\centering
	\includegraphics[width = 0.48\textwidth]{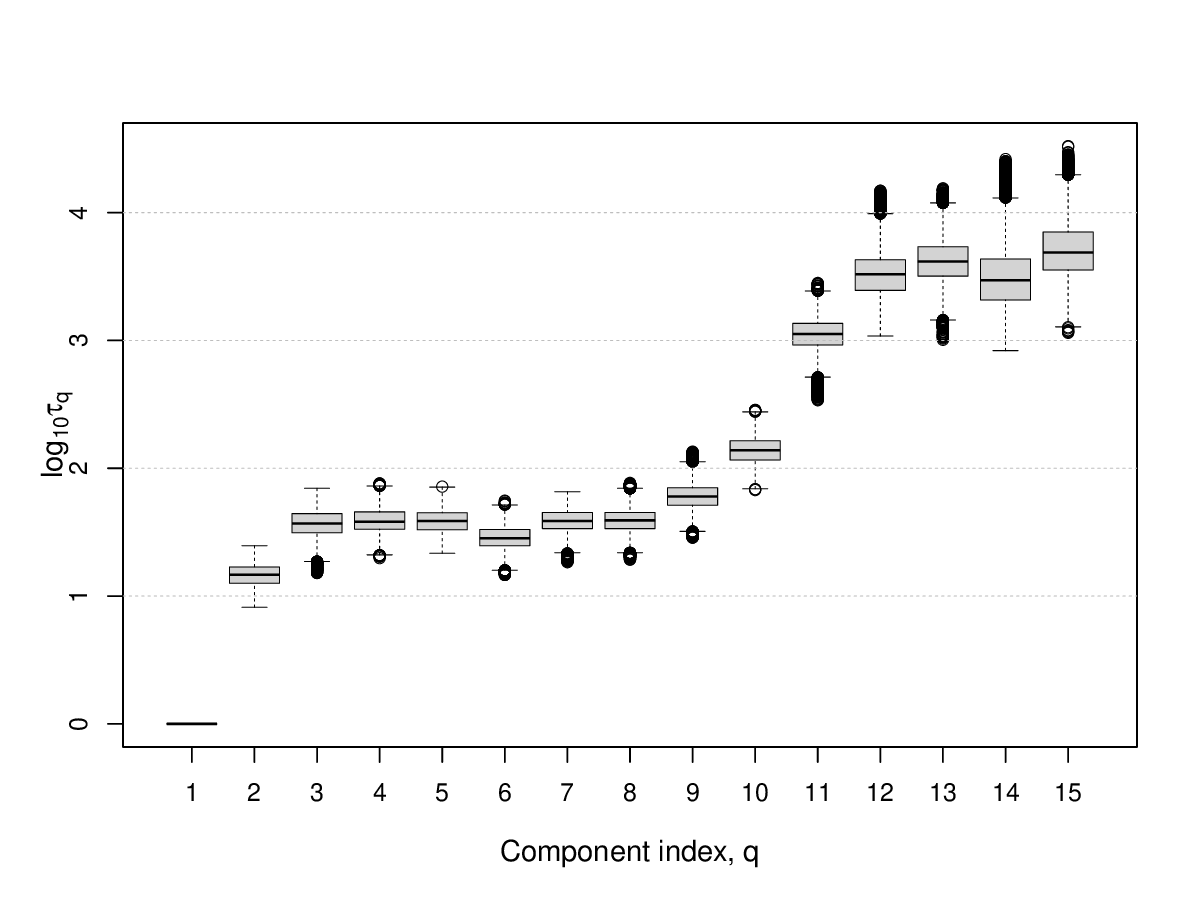}
    \includegraphics[width = 0.48\textwidth]{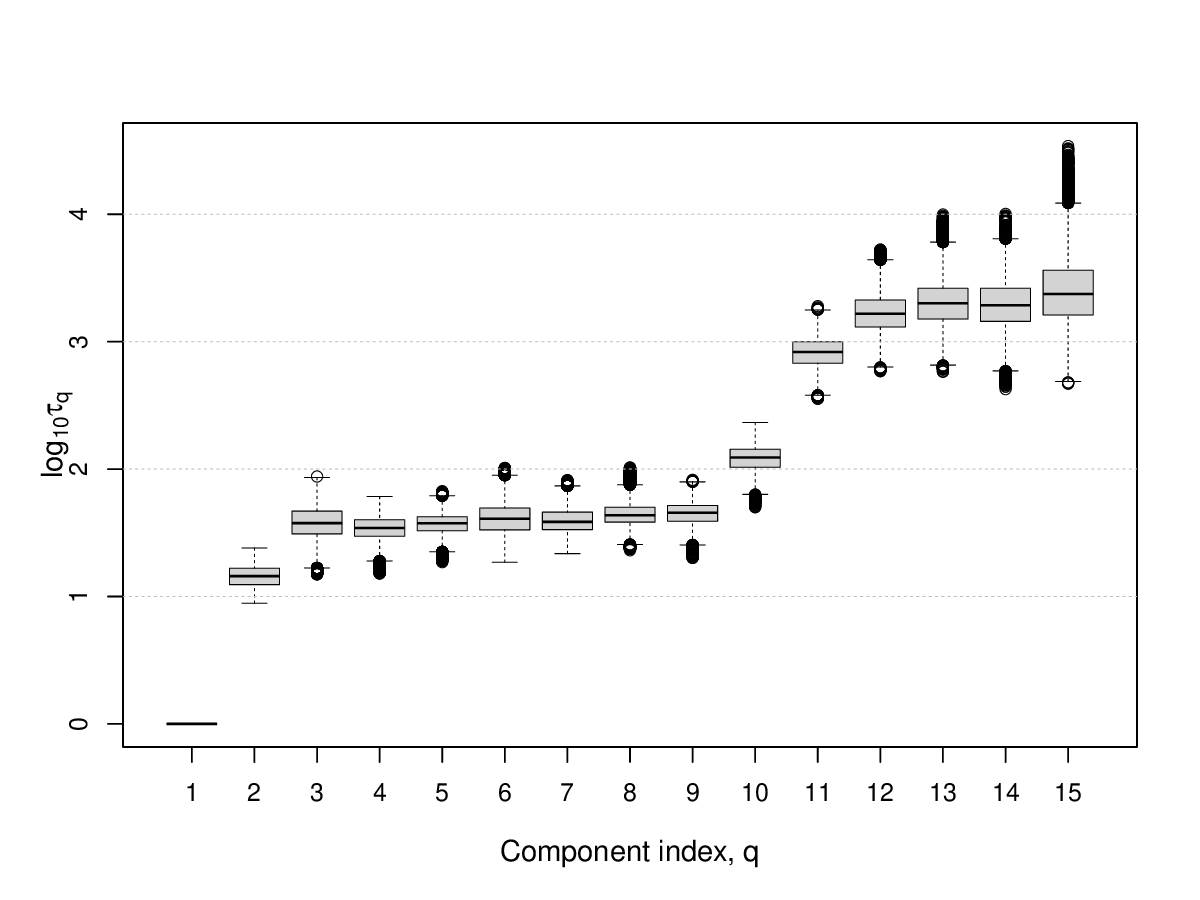}
	\caption{Posterior distributions of the shrinkage variables $\bmtau$ for the BPLS (Left) and L-BPLS (Right) models fitted to simulated data with using 10 latent components.}
	\label{fig:shrinkage}
\end{figure}

Figure \ref{fig:shrinkage} shows an example of the posterior distributions of $\bmtau$ elements when the proposed models are fitted to a synthetic dataset from Section 4 of the main paper; here, $N_{\mathrm{train}}=50$, $\ttP = 1000$ and $\sigma^2=\psi^2=0.5$. The true number of components used to generate the data was $\ttQ=10$ and the limit for the fitted number of components was $\ttQ^*=15$---the posterior variances of the loading matrix elements past the $10\Th$ column decrease by an order of magnitude which suggests the models can correctly identify the latent dimension. This kind of behaviour was consistent for choices of $N_{\mathrm{train}},~\ttP,~\sigma^2,$ and $\psi^2$.

\section{Posterior predictions from the model}\label{app:posterior_predictive}
\subsection{Estimating the moments of the posterior predictive} 
\label{app:estimating_the_moments_of_the_posterior_predictive}

An efficient alternative to using a Monte Carlo sample from the posterior predictive distribution of $\bfy|\bfx$ is to instead use the output of the Gibbs algorithm to obtain estimates of the response's expectation and variance, without simulating form the posterior predictive. These can be obtained from the $N_{\mathrm{mc}}$ posterior samples, $\left\{\Theta_m\right\}_{m=1}^{N_{\mathrm{mc}}}\sim\pi$, via Rao-Blackwellisation,
\begin{align*}
	\E[\bfy|\bfx] &= \E_\Theta\left[\E[\bfy|\Theta,\bfx]\right]\\
	&\approx \frac{1}{N_{\mathrm{mc}}}\sum_{m=1}^{N_{\mathrm{mc}}}\E[\bfy|\Theta_m,\bfx],\\
	\V[\bfy|\bfx] &=\E_\Theta\left[\V[\bfy|\Theta,\bfx]\right] + \V_\Theta\left[\E[\bfy|\Theta,\bfx]\right]\\
	&=\E_\Theta\left[\V[\bfy|\Theta,\bfx]\right] + \E_\Theta\left[\E[\bfy|\Theta,\bfx]\E[\bfy|\Theta,\bfx]^\top\right]-\E_\Theta\left[\E[\bfy|\Theta,\bfx]\right]\E_\Theta\left[\E[\bfy|\Theta,\bfx]\right]^\top\\
	&\approx\frac{1}{N_{\mathrm{mc}}}\left( \sum_{m=1}^{N_{\mathrm{mc}}}\V[\bfy|\Theta_m,\bfx]\right) + \frac{1}{N_{\mathrm{mc}}}\left(\sum_{m=1}^{N_{\mathrm{mc}}}\E[\bfy|\Theta_m,\bfx]\E[\bfy|\Theta_m,\bfx]^\top\right) \\ 
	&\qquad- \frac{1}{N_{\mathrm{mc}}^2}\left(\sum_{m=1}^{N_{\mathrm{mc}}}\E[\bfy|\Theta_m,\bfx]\right) \left(\sum_{n=1}^{N_{\mathrm{mc}}}\E[\bfy|\Theta_m,\bfx]\right)^\top.
\end{align*}

We note that $\E[\bfy|\Theta,\bfx] = CS_zW^\top\Sigma\inv\bfx$ leads with a matrix free of $\bfx$, thus the posterior predictive expectation estimate only requires a sufficient statistic which is the sum of those matrices. The same, however, cannot be obtained for the posterior predictive variance; whilst $\V[\bfy|\Theta,\bfx]$ is free of $\bfx$ (Section 3.5 of the main paper) the summands $\E[\bfy|\Theta,\bfx]\E[\bfy|\Theta,\bfx]^\top$ involve a $\bfx\bfx^\top$ term on the inside --- for a new test sample, the summation requires the storage of the whole (thinned) chain. 


\subsection{Predicting from the posterior mode}
\label{app:posterior_mode}
Throughout the main paper, we utilised the marginal posterior predictive distribution for producing point predictions of $\bfy$ given new $\bfx$ observations. Here we outline a comparison to predictions obtained based on the parameter posterior modes, denoted by $\Theta^m$. These were estimated from the MCMC output by identifying the sample with the highest posterior density value; this was an accurate approximation here and the results did not vary between different chains. We included two variants for each ss-BPLS and L-BPLS: (i) predictions derived conditional on a fixed $\widehat{\Theta^m}$, and (ii) predictions from new chains where $\widehat{(CB)^m}$ and $\widehat{C^m}$ respectively were held fixed at the mode and all other parameters were allowed to vary. The approaches were compared based on the relative predictive root mean-squared errors (RMSEs), where the prediction error of a modal model was divided by the error resulting from predicting from the full marginal posterior; for example, if all parameters were fixed at the mode this ratio would be $$\frac{\textsf{RMSE}\left(\E\left[\bfy_+ | \bfx_+ , \widehat{\Theta^m}\right]\right)}{\textsf{RMSE}\left(\E[\bfy_+ | \bfx_+ , \Dcal]\right)}.$$ The relative prediction accuracy results are given here in Table \ref{tab:model_comp} for the real data examples considered in the main paper and the results on synthetic datasets showed little to no change. On average the spike-and-slab approach saw mild improvements, particularly in the SERS data example, however, these results were subject to noise. The LASSO approach appears to suffer from predicting from the mode only, suggesting the combination of the Bayesian LASSO prior and the multiplicative shrinkage prior results in a marginal posterior predictive distribution reliable for prediction tasks. This can be linked to the results discussed in \cite{Rock2023} where desirable point predictions from the posterior predictive of a Bayesian LASSO model can be recovered by imposing mixtures of sparsity-inducing priors.

\begin{table}
\centering
\caption{Predictive strength of ss-BPLS and L-BPLS using a model with (i) all parameters $\Theta$ fixed at the posterior mode, and (ii) $CB$ and $C$ fixed at the posterior mode for ss-BPLS and L-BPLS respectively. The values are the ratios of the root-mean squared error (RMSE) for each modal model approach divided by the RMSE of the predictions made based on the marginal posterior predictive distribution of the responses for respective ss-BPLS and L-BPLS models.}\label{tab:model_comp}
\begin{tabular}{@{}ll|l|l|l|l@{}}
        &         & \multicolumn{2}{c|}{\textbf{ss-BPLS}} & \multicolumn{2}{c}{\textbf{L-BPLS}} \\\hline
\textbf{Dataset} & \textbf{Trait}   &   $\Theta$ fixed          &    $CB$ fixed         &  $\Theta$ fixed           &    $C$ fixed          \\ \hline
NIR-Grain   & Glucose &1.037~(0.088)  &0.994~(0.006)  &0.994~(0.069)  &1.000~(0.048)    \\
        & Ethanol & 0.979~(0.032)  &0.984~(0.044)  &1.474~(0.212)  &1.091~(0.042)   \\\hline
MIR-Milk    & HS      &  1.076~(0.038)  &1.007~(0.034)  &1.011~(0.031)  &1.010~(0.027)   \\
        & RCT     &  1.032~(0.081)  &0.998~(0.030)  &1.019~(0.038)  &1.013~(0.033)   \\
        & Casein  &0.996~(0.056)  &0.998~(0.028)  &1.012~(0.026)  &1.048~(0.052)\\ \hline
SERS-Milk   & pH      &0.908~(0.388)  &0.909~(0.285)  &1.330~(0.760)  &1.487~(0.452)  
\end{tabular}
\end{table}
\section{Additional simulation results}
\label{app:sim_results}

\subsection{Sparse loading matrix}
\label{app:sim_sparse}

We consider two types of sparsity: column-wise, and element-wise. In the former, we set the odd-numbered columns of $C$ to zero, and in the latter we randomly set half of the entries in each row to zero. For consistent comparisons with the previous set of experiments, the non-zero entries of $C$ must be scaled to preserve the signal-to-noise ratio for given $y^r_n$ element of $\bfy_n$ ($r=1,...,\ttR; n=1,...,N$) which we define to be 
\begin{align*}
    \mathsf{RATIO}_r:=&  \frac{\V\left[\sum_{q =1}^{\ttQ} c_{rq}z^q_n\right]}{ \V[\eta_n^r]}\\
    =&\frac{\sum_{q =1}^{\ttQ} \V\left[c_{rq}z^q_n\right]}{ \psi^2}\\
    =&\frac{\sum_{q =1}^{\ttQ} \V\left[c_{rq}\right]}{ \psi^2},
\end{align*}
where the third equality follows from both $c_{rq}$ and $z_n^r$ being zero-mean, independent and $\V[z_n^q]=1$; $\V[c_{rq}z^q_n] = \E[c_{rq}^2]\E\left[(z^q_n)^2\right] =\V[c_{rq}]$. The synthetic datasets in Section 4.1 of the paper were generated with $\mathsf{RATIO}_r = \psi^{-2}$, $r=1,...,\ttR$---here, scaling of the entries by $\sqrt{2}$ results in $\sum_{q =1}^{\ttQ} \V\left[c_{rq}\right] = (\ttQ/2)\times(2/\ttQ) = 1$ which preserves the signal-to-noise ratio.

\begin{figure}[tb]
	\centering
	\includegraphics[width = \textwidth]{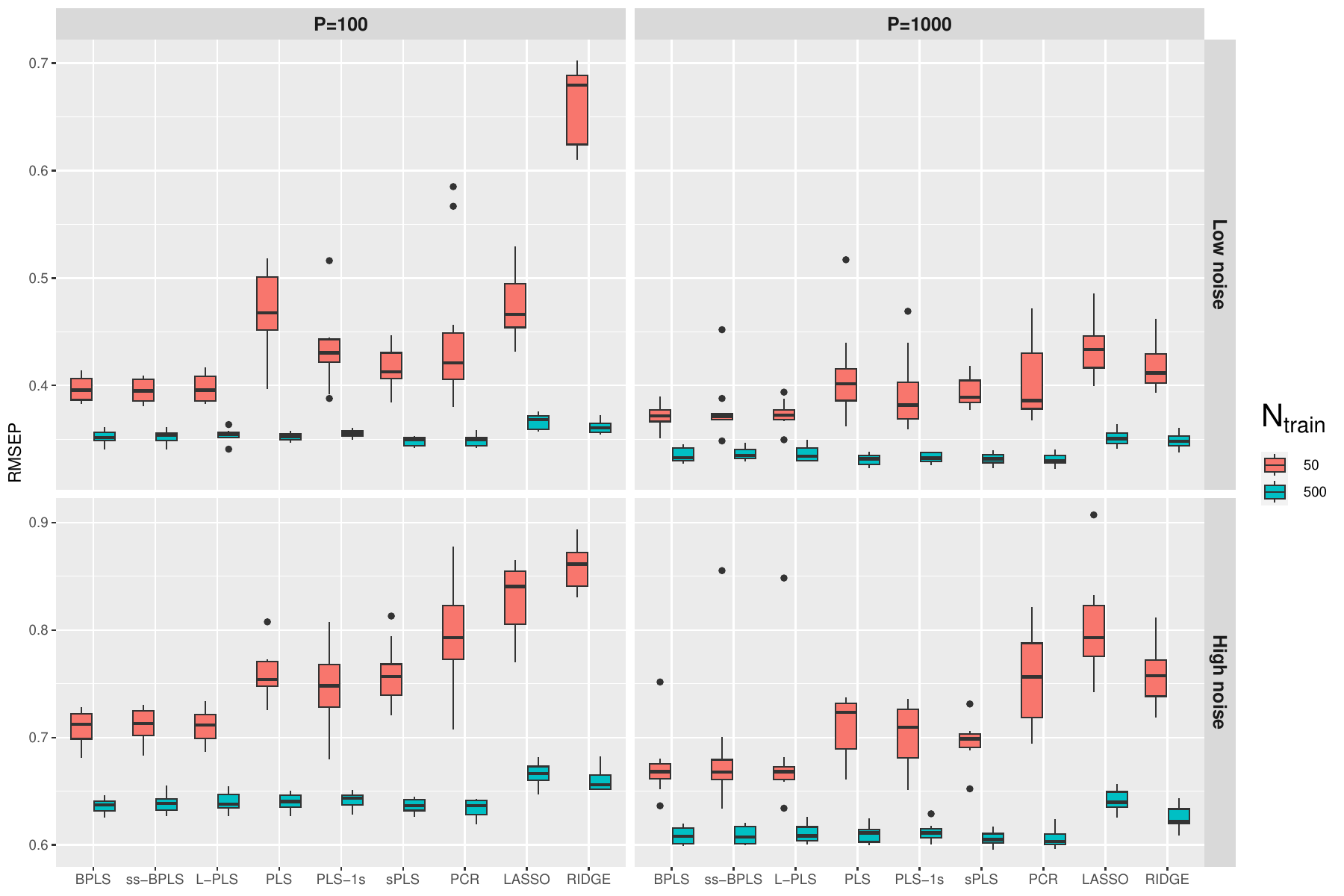}
	\caption{RMSEP on simulated datasets with column-wise sparsity in the $C$ matrix. Low-noise and high-noise scenarios correspond to $\sigma^2=\psi^2=0.1$ and $\sigma^2=\psi^2=0.5$ respectively.}
	\label{fig:simboxplotfreess}
\end{figure}

\begin{figure}[tb]
	\centering
	\includegraphics[width = \textwidth]{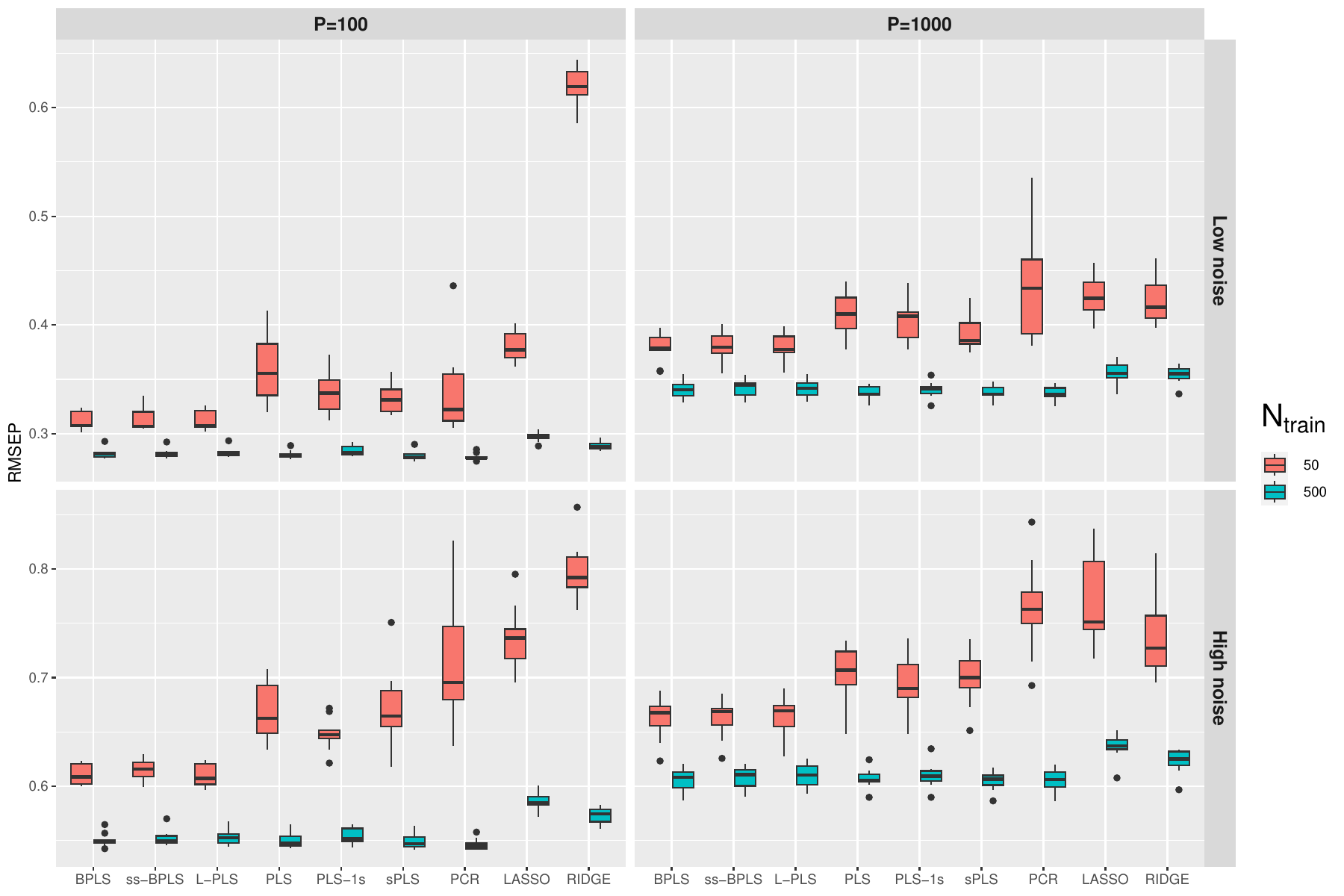}
	\caption{RMSEP on simulated datasets with element-wise sparsity in the $C$ matrix. Low-noise and high-noise scenarios correspond to $\sigma^2=\psi^2=0.1$ and $\sigma^2=\psi^2=0.5$ respectively.}
	\label{fig:simboxplotfreelasso}
\end{figure}

Recall that we varied the following: $N\in\{50,500\},~\ttP\in\{100,1000\},~\sigma^2\in\{0.1,0.5\}$, with $\ttQ=10,~\ttR=4$ and $\psi^2 = \sigma^2$. Figures \ref{fig:simboxplotfreess} and \ref{fig:simboxplotfreelasso} show the results of comparisons on different data-generating mechanisms under different scenarios. The BPLS methods all perform reliably well compared to the competitors. In fact, even the standard BPLS model without explicit sparsity is robust here suggesting that the model fit favors a latent space rotation which leads to overall parsimony. It should, however, be noted that the sparse models improve the point predictions on real datasets as per the results in Section 5 of the main paper. 

\section{Additional data results}
\label{app:additional_data}

\begin{figure}[tb]
	\centering
	\includegraphics[width = 0.8\textwidth]{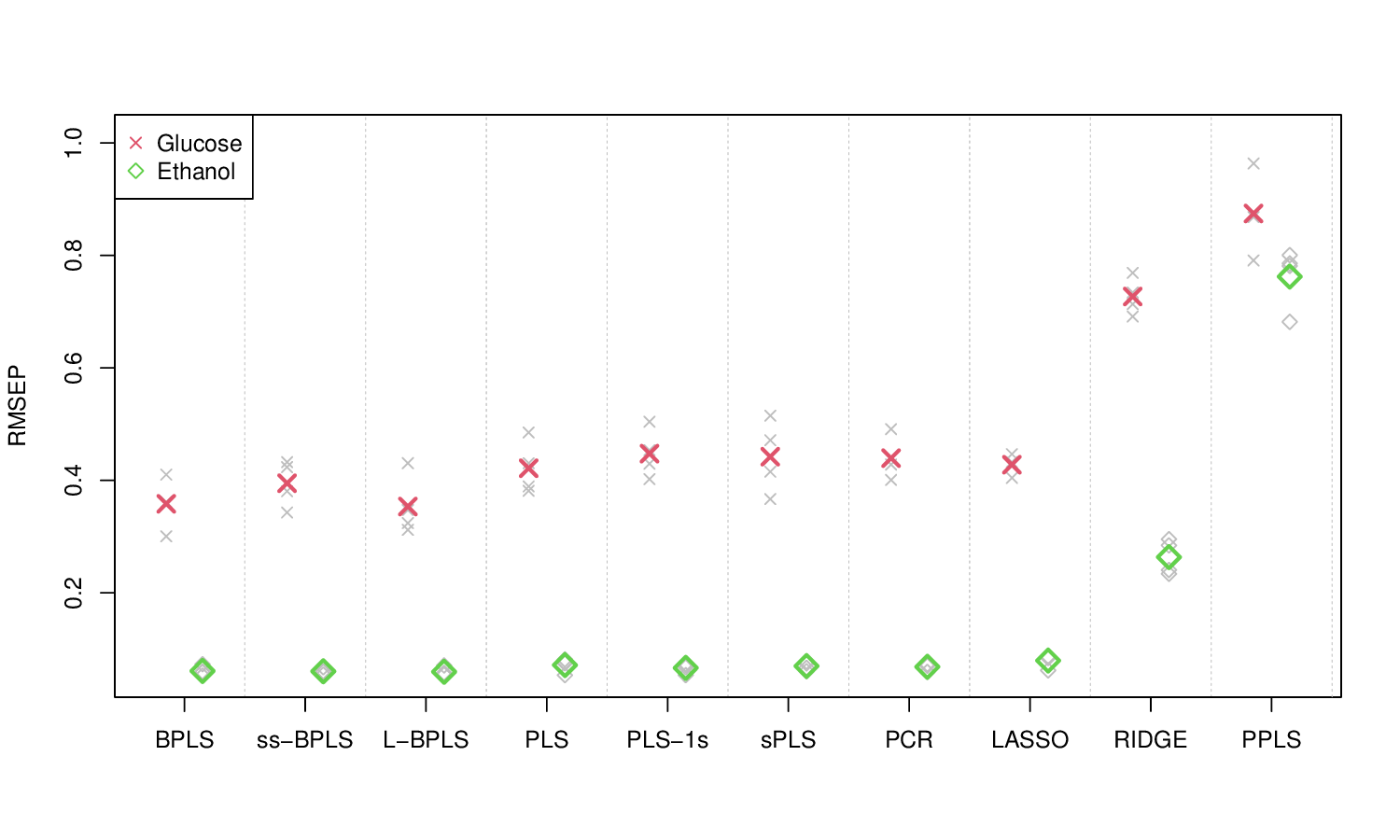}
	\caption{Comparison of prediction methods on the grain mash NIR spectral data where accuracy is reported as the root-mean squared error on the prediction data (RMSEP) for each response component. Grey symbols mark the RMSEP values per data fold and coloured symbols give their averages.}
	\label{fig:NIR_grain3x1}
\end{figure}

\begin{figure}[tb]
	\centering
	\includegraphics[width = 0.8\textwidth]{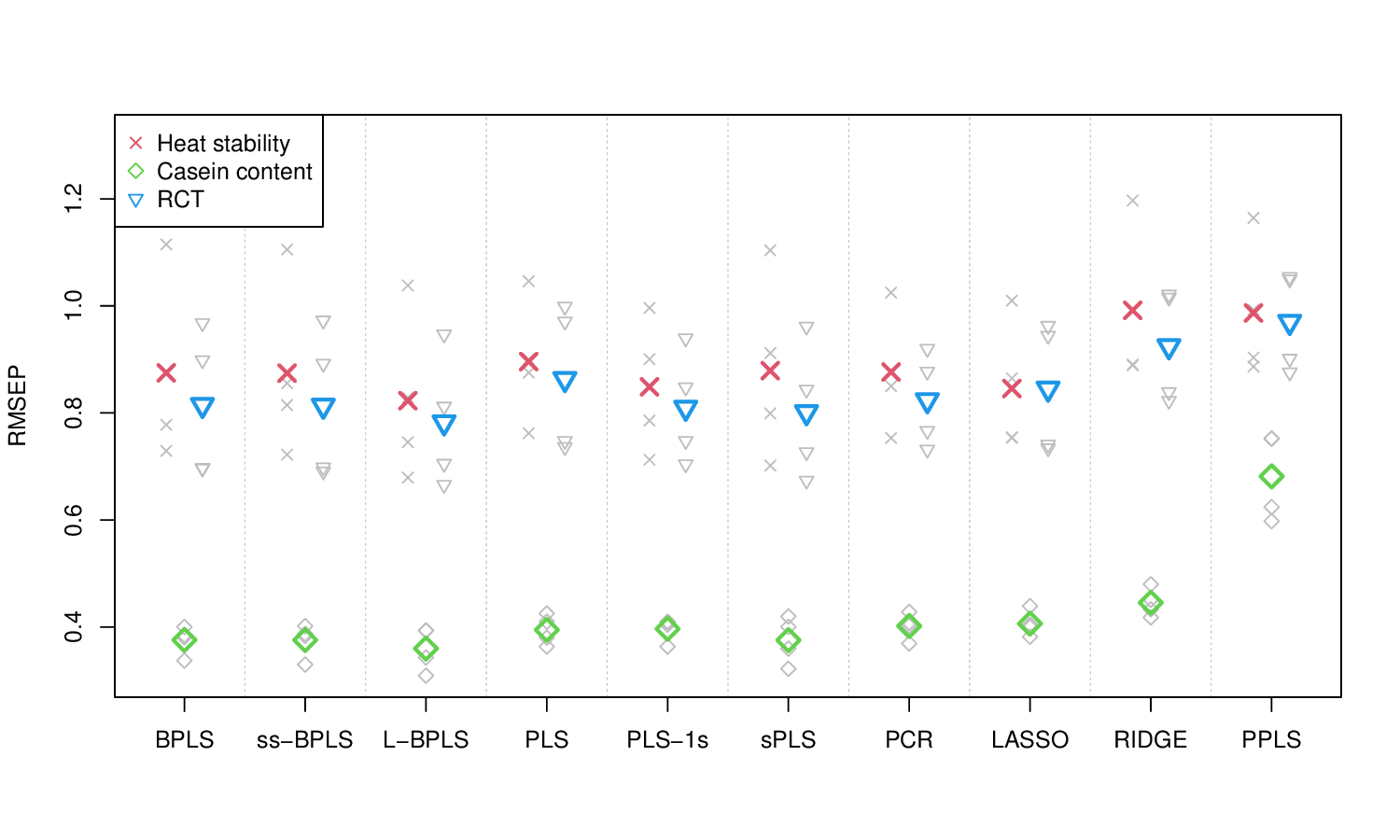}
	\caption{Comparison of prediction methods on the milk sample MIR spectral data where accuracy is reported as the root-mean squared error on the prediction data (RMSEP) for each response component. Grey symbols mark the RMSEP values per data fold and coloured symbols give their averages.}
	\label{fig:MIR_milk3x1}
\end{figure}

\begin{figure}[tb]
	\centering
	\includegraphics[width = 0.66\textwidth]{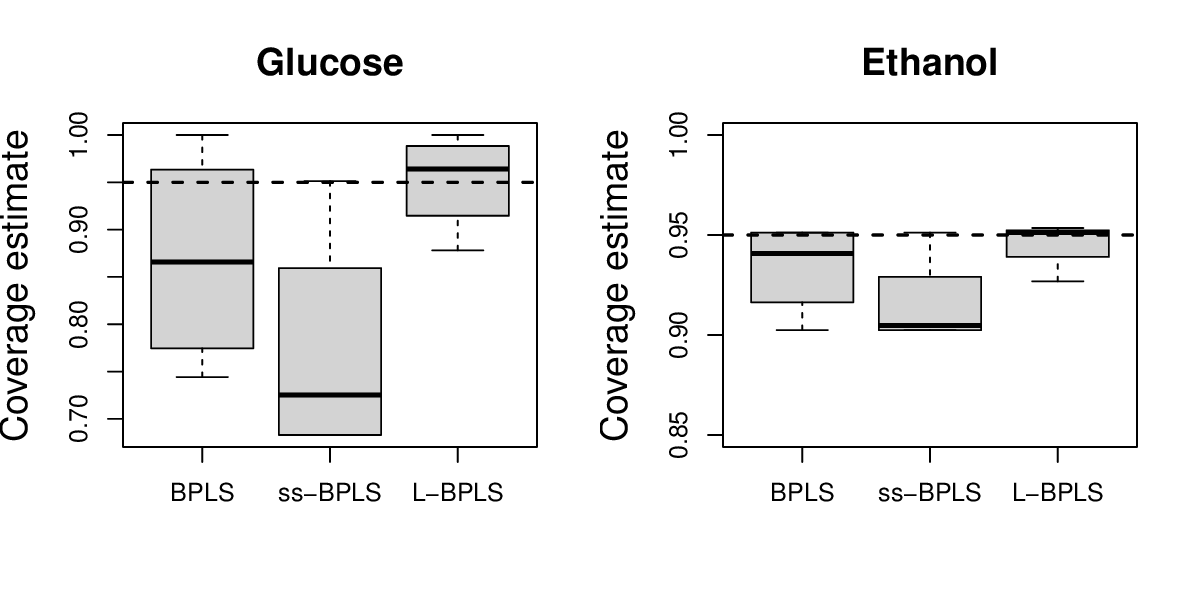}
	\caption{Estimates of $95\%$ prediction interval coverage in the grain mash content prediction from NIR spectra example. Estimates are based on the proportion of test samples contained within the predicted intervals.}
	\label{fig:NIRgrain_coverage3x1}
\end{figure}

Figures \ref{fig:NIR_grain3x1} and \ref{fig:MIR_milk3x1} illustrate the prediction accuracy of BPLS models and its competitors in various instances; the figures correspond directly to Tables 1 and 2  in the main paper. Additionally, Figure \ref{fig:NIRgrain_coverage3x1} reports the coverage estimates of these intervals; caution should be taken when interpreting these as in each instance the test set contained fewer than 60 samples.



\subsection{Comparison to Bayesian factor analysis}\label{app:bfa}
The BPLS regression model formulation can be seen as a special case of a Bayesian factor analysis (BFA) model with a block structure. Here, we provide a comparison between the proposed BPLS approach and standard BFA models utilising the multiplicative gamma process prior. The first approach is a direct implementation of the model detailed in \cite{BhDu2011}; we have implemented the model exactly as described therein and below refer to this approach as BFA-1. The BFA-1 approach differs from the proposed BPLS approach as in BFA-1 the shrinkage hyperparameters are included in the inference \citep[Step 6 in Section 3.1 of][]{BhDu2011} and very vague priors are imposed on all uniqueness variables \citep[Section 4.1 of][]{BhDu2011}. With the BFA-1 approach we experienced convergence issues with respect to the shrinkage hyperparameters, generally requiring many more iterations to get an acceptable effective sample size. We also considered two further BFA models where the shrinkage hyperparameters are fixed but two different uniqueness priors are considered:
\begin{itemize}
    \item BFA-2 \textit{(common uninformative prior)}: $\sigma_p^{-2}\sim\gamdist(2.5,0.1),~p=1,\ldots,\ttP$, and $\psi_r^{-2}\sim\gamdist(2.5,0.1),~r=1,\ldots,\ttR$; and
    \item BFA-3 \textit{(common empirical prior)}: $\sigma_p^{-2}\sim\gamdist(2.5,B_\sigma^p),~p=1,\ldots,\ttP$, and $\psi_r^{-2}\sim\gamdist(2.5,B_\psi^r),~r=1,\ldots,\ttR$, with $B_\sigma^p= (2.5-1)/\left[V^{-1}\right]_{pp}$, where $V^{-1}$ is a ridge-type estimator of the sample precision matrix of $\bfX$\footnote{$V^{-1} = (b_0+N/2)\left(b_0 I_\ttP + 0.5 \sum_{n=1}^N\bfx_n\bfx_n^\top\right)^{-1}$ and $b_0 = 0.5$; the results were similar for other choices of $b_0$.}; a similar definition follows for $B_\psi^r,~r=1,...\ttR$.
\end{itemize}

Table \ref{tab:bfa_comp} gives the prediction performance of the BFA-1, BFA-2, BFA-3, BPLS and L-BPLS for the three real datasets considered as well as for the most challenging synthetic dataset, i.e.\ $N=50$ samples, $\ttP=1000$ variables, high noise.  Results demonstrate that the BPLS approaches perform strongly, with the L-BPLS approach providing the smallest root-mean squared error of predictions in the majority of cases. Note that the synthetic data were simulated from a model where all uniqueness values were the same; the BFA models, with common uniqueness priors for all variables, would most closely resemble this scenario while observed spectral data typically have notable variation in uniqueness estimates. We experienced substantial issues when inferring the shrinkage in the BFA-1 model applied to real spectral data. In all three BFA cases, the posterior of the hyperparameter $\alpha$ had substantial mass below 2 which, as noted in \cite{Dura2017}, does not impose shrinkage on the latent variables.

\begin{table}[h]
\centering
\caption{The root-mean squared error of predictions for the BFA-1, BFA-2 and BFA-3 approaches, and for the 
two BPLS models.
$^*$Markov chain corresponding to shrinkage hyperparameters failed to converge within 50,000 iterations.}\label{tab:bfa_comp}
\begin{tabular}{@{}ll|l|l|l|l|l@{}}
        &         & \multicolumn{3}{c|}{\textbf{Bayesian factor analysis}} & \multicolumn{2}{c}{\textbf{Bayesian PLS}} \\\hline
\textbf{Dataset} & \textbf{Trait}   &   BFA-1 (B\&D)         &    BFA-2   &BFA-3      &  BPLS           &    L-BPLS          \\ \hline
NIR-Grain   & Glucose &0.405 (0.082)&0.382 (0.043) &0.439 (0.090)  &0.358 (0.045)  &\textbf{0.353} (0.053)    \\
        & Ethanol &0.064 (0.007)& 0.064 (0.006) &0.076 (0.016)  &0.061 (0.009)  &\textbf{0.060} (0.008)  \\\hline
MIR-Milk    & HS      &0.899 (0.117)&  0.875 (0.161)  &0.901 (0.103)  &0.875 (0.172)  &\textbf{0.823} (0.156)   \\
        & RCT     &0.833 (0.122)&  0.825 (0.128)  &0.855 (0.129)  &0.815 (0.139)  &\textbf{0.782} (0.126)   \\
        & Casein  &0.377 (0.029)&0.366 (0.037)  &0.379 (0.026)  &0.376 (0.027)  &\textbf{0.360} (0.041)\\ \hline
SERS-Milk   & pH      &$^*$0.575 (0.340)&0.640 (0.442)  &\textbf{0.574} (0.264)  & 0.627 (0.248)  & 0.614 (0.232)\\
\hline
Synthetic && 0.658 (0.009)&\textbf{0.658} (0.008) & 0.658 (0.009) &0.666 (0.030) &0.675 (0.064)\\\hline
\end{tabular}
\end{table}

\subsection{Smaller training datasets}
\label{app:small_data}
\begin{figure}[tb]
	\centering
	\includegraphics[width = 0.8\textwidth]{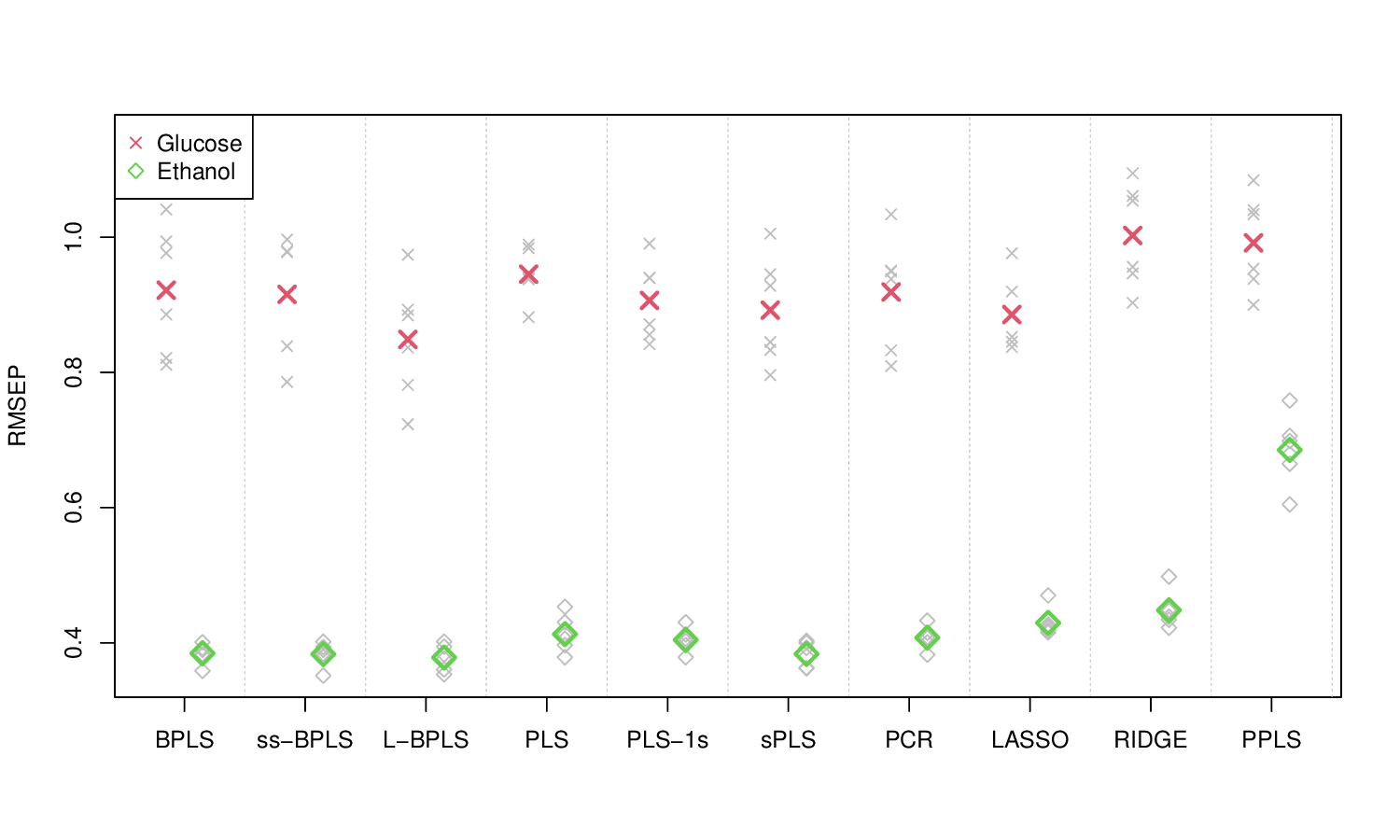}
	\caption{Comparison of prediction methods' performance on the grain mash NIR spectral data where accuracy is reported as the RMSEP for each response component. Here, each instance involved fitting the models to two quarters of data with predictions on the remaining two quarters; there were six instances in total. Grey symbols mark the RMSEP values per data fold and coloured symbols give their averages.}
	\label{fig:NIR_grain2x2}
\end{figure}

\begin{figure}[tb]
	\centering
	\includegraphics[width = 0.8\textwidth]{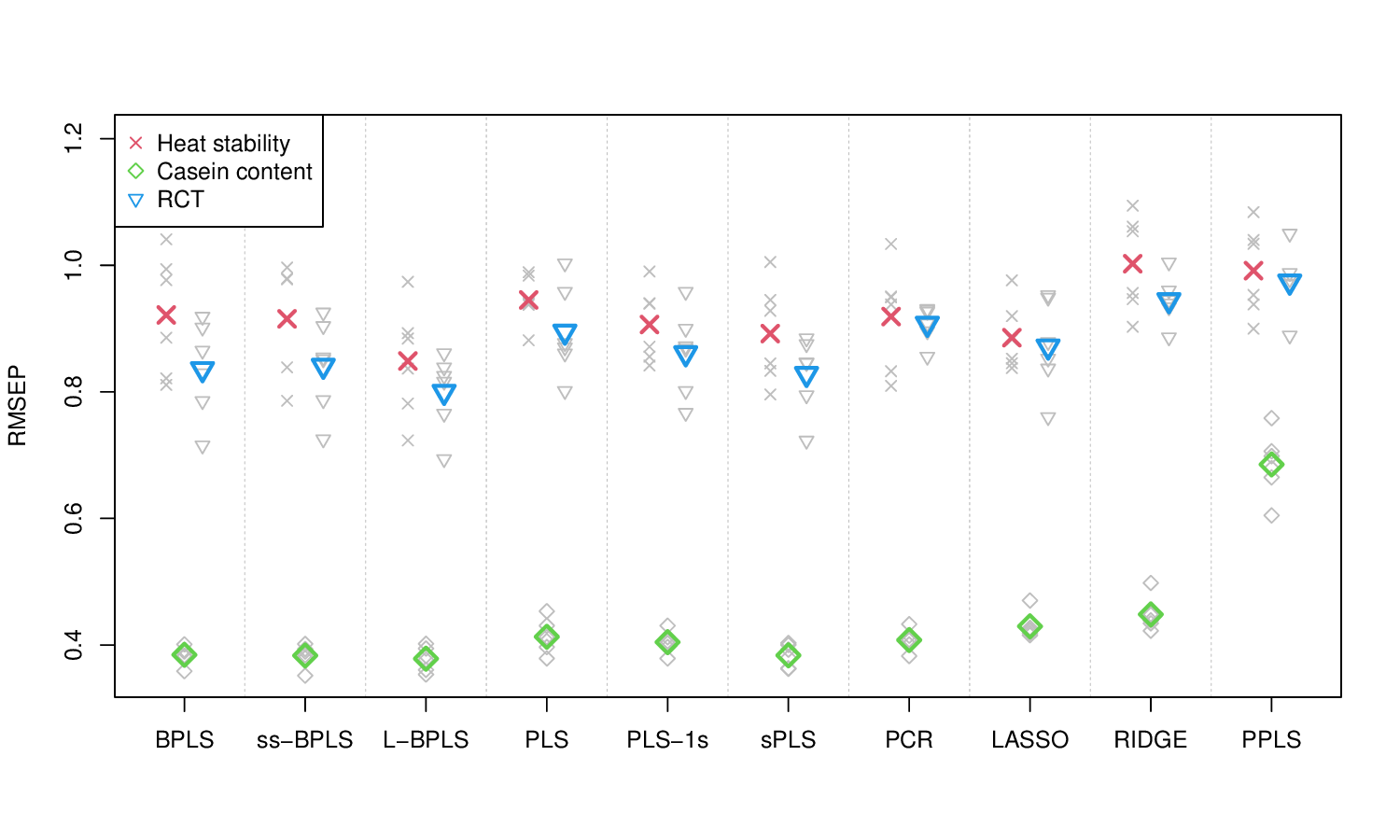}
	\caption{Comparison of prediction methods' performance on the milk sample MIR spectral data where accuracy is reported as the RMSEP for each response component. Here, each instance involved fitting the models to two quarters of data with predictions on the remaining two quarters; there were six instances in total. Grey symbols mark the RMSEP values per data fold and coloured symbols give their averages.}
	\label{fig:MIR_milk2x2}
\end{figure}

\begin{figure}[tb]
	\centering
	\includegraphics[width = 0.66\textwidth]{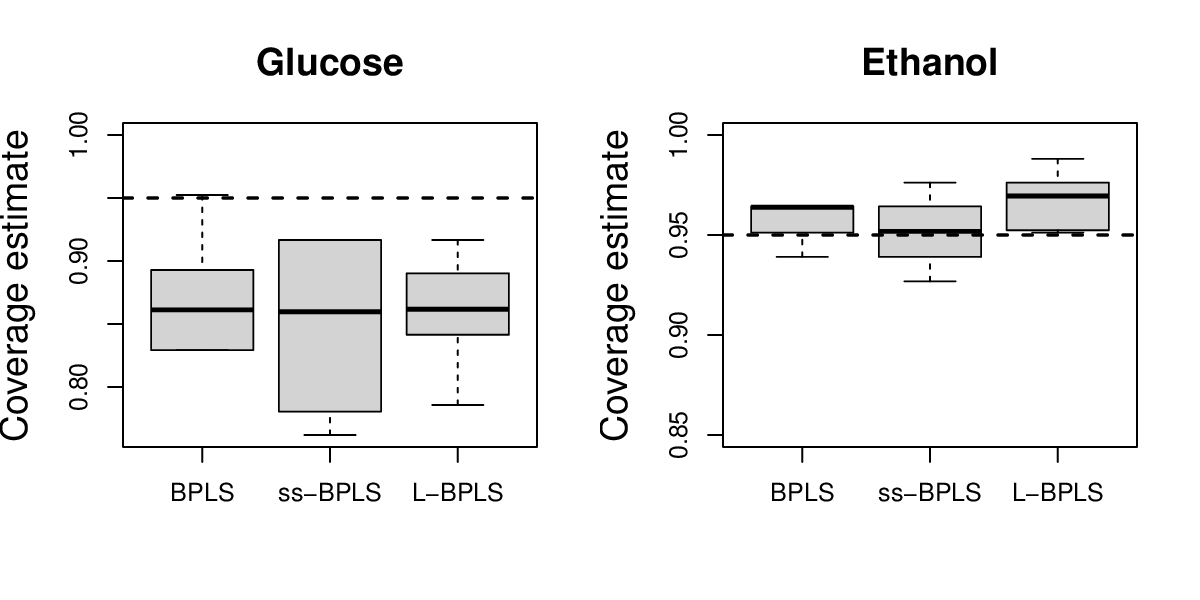}
	\caption{Estimates of $95\%$ prediction interval coverage in the grain mash content prediction from NIR spectral data. Estimates are based on the proportion of test samples contained within the predicted intervals. Here, each instance involved fitting the models to two quarters of data with predictions on the remaining two quarters; there were six instances in total.}
	\label{fig:NIRgrain_coverage2x2}
\end{figure}

\begin{figure}[tb]
	\centering
	\includegraphics[width = \textwidth]{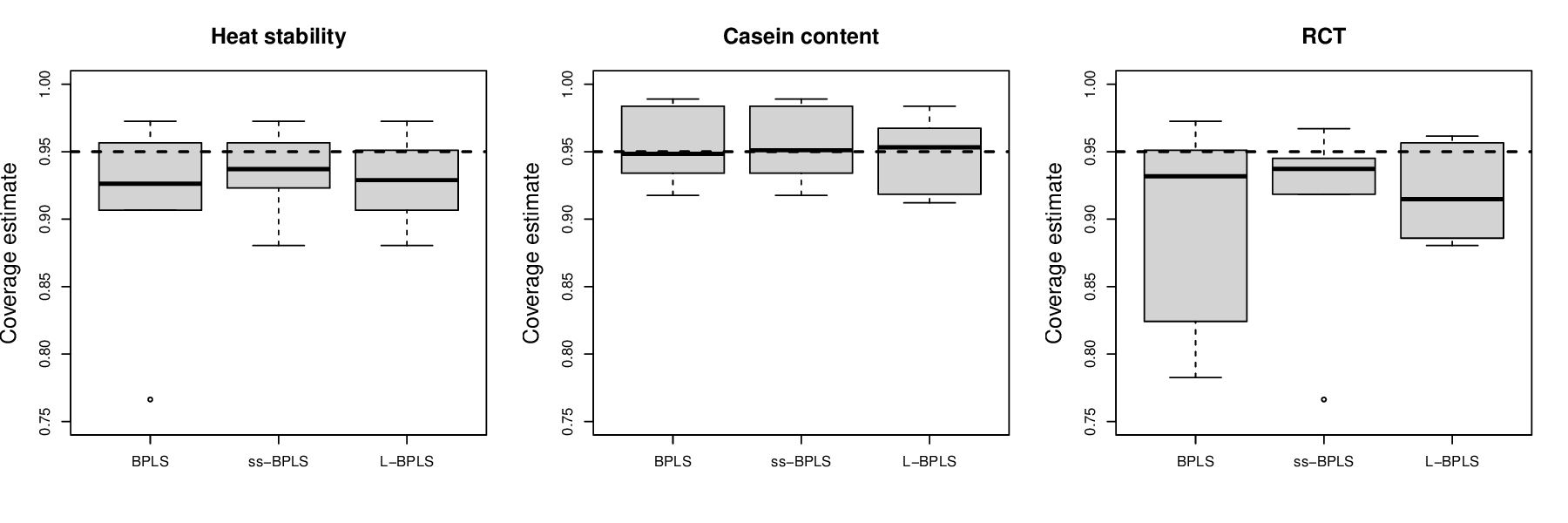}
	\caption{Estimates of $95\%$ prediction interval coverage in the milk trait prediction from MIR spectral data. Estimates are based on the proportion of test samples contained within the predicted intervals. Here, each instance involved fitting the models to two quarters of data with predictions on the remaining two quarters; there were six instances in total.}
	\label{fig:MIRmilk_coverage2x2}
\end{figure}

Here, we illustrate the performance of the BPLS and competitor methods on the NIR and MIR datasets under different training-test data splits; the SERS dataset is omitted due to its small $N$. As in Sections 4.2 and 5 of the main paper, each dataset is split into four folds but now we fit the models to only two folds and predict on the remaining two; in total there are six possible combinations. Tables \ref{tab:nir_grain2x2} and \ref{tab:mir_milk2x2}, and Figures \ref{fig:NIR_grain2x2} and \ref{fig:MIR_milk2x2} show the results of the analyses which show similar patterns to results obtained when models were trained on three folds. Furthermore, we provide estimates of the coverage in Figures \ref{fig:NIRgrain_coverage2x2} and \ref{fig:MIRmilk_coverage2x2}.


\begin{table}[tb]
	\caption{Comparison of methods' performance on the grain mash NIR spectral data where half the data were used for training and the remaining half was used a test set. Estimates  are given as `mean (standard deviation)' over six validation combinations, with bold font indicating optimal performance. $^*$Due to numerical instability, $\widehat{\ttQ}_y$ was picked manually.}
	\label{tab:nir_grain2x2}
	\centering

	\begin{tabular}{l|cc|cc}

	\hline
	&\multicolumn{2}{c}{\textbf{Glucose}} & \multicolumn{2}{|c}{\textbf{Ethanol}} \\
	&RMSEP&$\widehat{\ttQ}_y$& RMSEP&$\widehat{\ttQ}_y$  \\
	\hline
BPLS & 0.443 (0.132) & 7.3 (1.0) & 0.070 (0.014) & 5.3 (0.8) \\ 
  ss-BPLS & 0.431 (0.128) & 7.2 (1.2) & \textbf{0.066} (0.013) & 5.5 (1.0) \\ 
  L-BPLS & \textbf{0.429} (0.126) & 11.7 (3.0) & 0.066 (0.017) & 6.3 (1.2) \\ \hline
  PLS & 0.603 (0.093) & 5.8 (2.5) & 0.086 (0.024) & 8.2 (1.7) \\ 
  PLS-1s & 0.499 (0.076) & 8.5 (2.1) & 0.080 (0.013) & 10.3 (2.9) \\ 
  sPLS & 0.528 (0.137) & 15.5 (4.5) & 0.083 (0.013) & 12.5 (5.1) \\ 
  PCR & 0.629 (0.088) & 9.3 (5.2) & 0.083 (0.009) & 19.0 (2.1) \\ 
  LASSO & 0.487 (0.048) & -- & 0.084 (0.006) & -- \\ 
  RIDGE & 0.756 (0.043) &-- & 0.269 (0.020) & -- \\ 
  PPLS & 0.880 (0.046) & $^*$1.0 (0.0) & 0.765 (0.046) & $^*$1.0 (0.0)\\
	\hline

	\end{tabular}
\end{table}

\begin{table}[tb]
	\caption{Comparison of methods' performance on the milk MIR spectral data where half the data were used for training and the remaining half was used a test set. Estimates are given as `mean (standard deviation)' over six validation combinations, with bold font indicating optimal performance. $^*$Due to numerical instability, $\widehat{\ttQ}_y$  was picked manually.}
	\label{tab:mir_milk2x2}
	\centering

	\begin{tabular}{l|cc|cc|cc}

	\hline
	& \multicolumn{2}{c}{\textbf{Heat stability}} &  \multicolumn{2}{|c}{\textbf{Casein content}}&\multicolumn{2}{|c}{\textbf{RCT}}   \\
	& RMSEP&$\widehat{\ttQ}_y$ &RMSEP&$\widehat{\ttQ}_y$& RMSEP&$\widehat{\ttQ}_y$ \\
	\hline
BPLS & 0.922 (0.096) & 6.2 (2.2) & 0.384 (0.014) & 5.3 (0.8) & 0.836 (0.076) & 5.7 (2.0) \\ 
  ss-BPLS & 0.915 (0.086) & 5.5 (2.1) & 0.383 (0.017) & 5.0 (0.9) & 0.841 (0.074) & 5.2 (0.8) \\ 
  L-BPLS & \textbf{0.849} (0.089) & 16.7 (1.6) & \textbf{0.378} (0.019) & 14.7 (3.3) & \textbf{0.800} (0.061) & 16.3 (2.3) \\ \hline
  PLS & 0.945 (0.039) & 4.0 (1.9) & 0.413 (0.026) & 3.8 (0.4) & 0.895 (0.073) & 4.7 (2.3) \\ 
  PLS-1s & 0.906 (0.059) & 5.8 (0.8) & 0.404 (0.017) & 4.0 (0.0) & 0.861 (0.069) & 5.7 (2.0) \\ 
  sPLS & 0.892 (0.080) & 12.5 (2.6) & 0.384 (0.018) & 9.2 (3.3) & 0.828 (0.060) & 10.7 (4.4) \\ 
  PCR & 0.919 (0.084) & 4.5 (1.0) & 0.408 (0.017) & 4.3 (0.5) & 0.908 (0.029) & 4.0 (2.8) \\ 
  LASSO & 0.886 (0.054) & -- & 0.430 (0.020) & -- & 0.872 (0.073) & -- \\ 
  RIDGE & 1.002 (0.077) & -- & 0.448 (0.026) &-- & 0.945 (0.038) & -- \\ 
  PPLS &0.992 (0.071) & $^*$2.0 (0.0) & 0.685 (0.051) & $^*$2.0 (0.0) & 0.974 (0.051) & $^*$2.0 (0.0)\\
	\hline

	\end{tabular}
\end{table}

\subsection{Multivariate response modelling}
\label{app:multivariate}

\begin{figure}[tb]
	\centering
	\includegraphics[width = 0.45\textwidth]{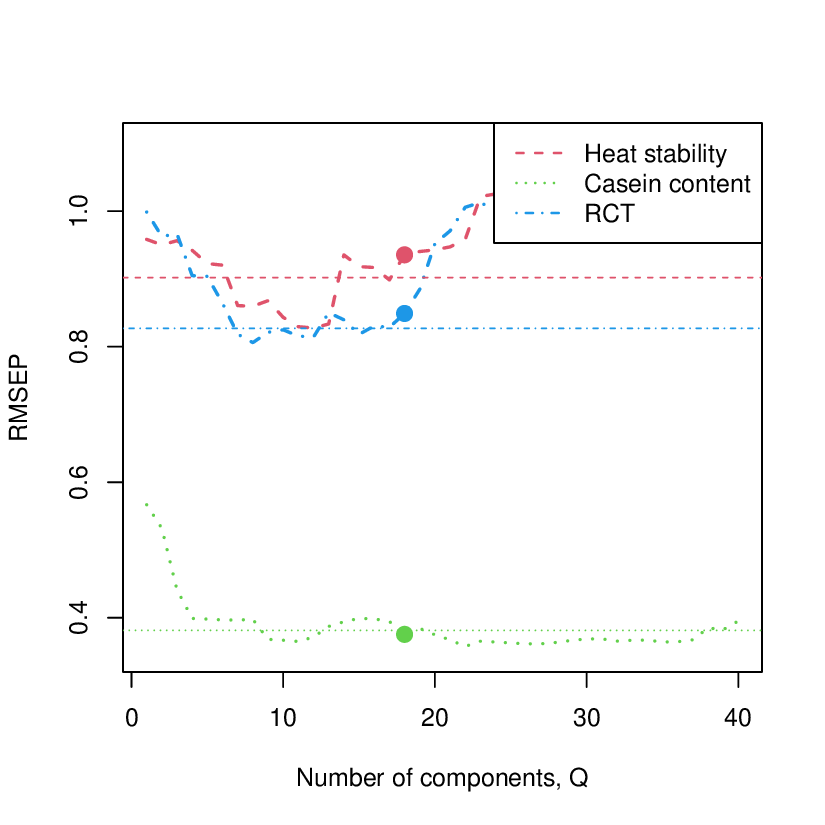}
	\includegraphics[width = 0.45\textwidth]{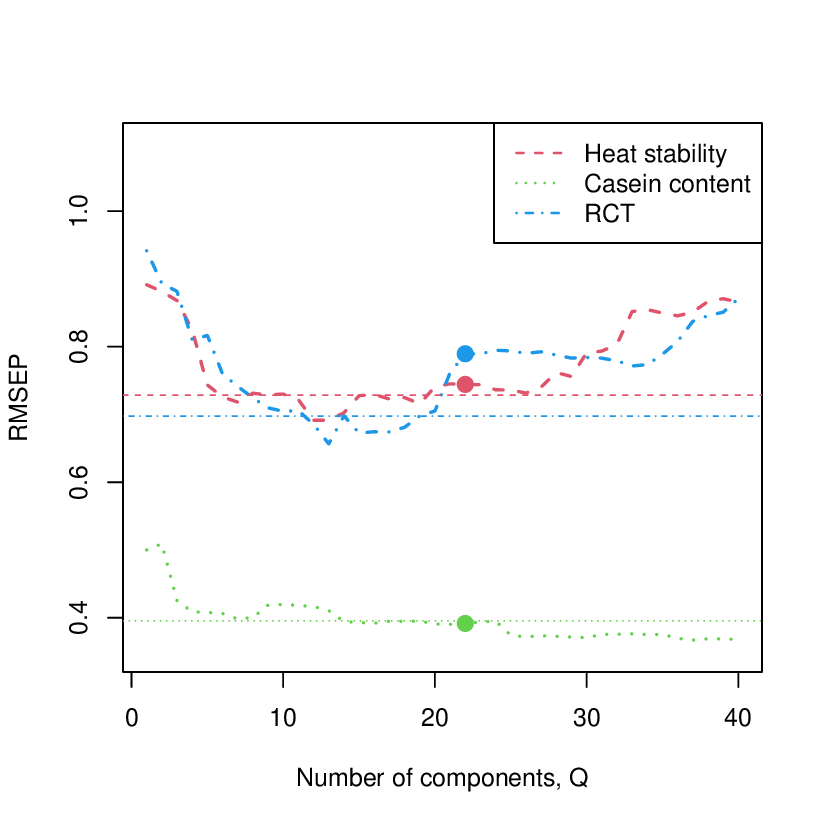}
	\caption{Root mean squared error of predictions from multivariate PLS fits to the milk MIR spectra with different numbers of components $\ttQ$. The $\newmoon$ symbols show the optimal $\ttQ$ from cross-validation based on the average RMSE across the three responses. Horizontal lines show the RMSEP values from the single fit of the BPLS model. (Left: $N_{\mathrm{train}} = 180$. Right: $N_{\mathrm{train}} = 270$.)}
	\label{fig:pls_MV}
\end{figure}

We wish emphasize that the BPLS models produce predictions for multivariate responses. In Figure \ref{fig:pls_MV}, we compare the standard BPLS model to a multivariate fit of PLS in one instance of the data split; the multivariate PLS model was chosen based on the unweighted average mean-squared error among the response components. Figure \ref{fig:pls_MV} illustrates how here this approach can overestimate the number of components resulting in overfitting.

\end{document}